\documentclass[journal]{IEEEtran}
\usepackage[T1]{fontenc}

\usepackage{array}

\usepackage{graphicx} 
\graphicspath{{./figs/}}

\usepackage{subcaption}
\usepackage{amsmath, amssymb, amsfonts, amsthm}
\usepackage{color}
\usepackage{mathtools}
\usepackage{algorithm}
\usepackage[noend]{algorithmic}
\usepackage{cuted}

\usepackage{multirow}
\usepackage{float}
\usepackage{url} 
\usepackage{comment}
\usepackage{cite}

\usepackage{lipsum}

\usepackage[acronym]{glossaries}
\newacronym{tx}{Tx}{transmit}
\newacronym{rx}{Rx}{receiver}
\newacronym{mimo}{MIMO}{multiple-input multiple-output}
\newacronym{siso}{SISO}{single-input single-output}
\newacronym{simo}{SIMO}{single-input multiple-output}
\newacronym{fir}{FIR}{finite impulse response}

\newacronym{mumimo}{MU-MIMO}{multi-user MIMO}
\newacronym{ula}{ULA}{uniform linear array}
\newacronym{fista}{FISTA}{Fast Iterative Shrinkage-Thresholding Algorithm}
\newacronym{pg}{PG}{projected gradient}
\newacronym{apg}{APG}{accelerated PG}
\newacronym{am}{AM}{alternating minimization}
\newacronym{rcs}{RCS}{radar cross-section}
\newacronym{isac}{ISAC}{integrated sensing and communication}
\newacronym{psl}{PSL}{peak sidelobe level}
\newacronym{isl}{ISL}{integrated sidelobe level}

\newacronym{bpm}{BPM}{beampattern matching}
\newacronym{svd}{SVD}{singular value decomposition}

\newacronym{snr}{SNR}{signal-to-noise ratio}
\newacronym{sinr}{SINR}{signal-to-interference-plus-noise ratio}

\newacronym{mui}{MUI}{multi-user interference}
\newacronym{pa}{PA}{power amplifier}
\newacronym{rf}{RF}{radio-frequency}
\newacronym{par}{PAPR}{peak-to-average-power-ratio}
\newacronym{cm}{CM}{constant-modulus}
\newacronym{qpsk}{QPSK}{quadrature phase shift keying}

\newacronym{mse}{MSE}{mean square error}
\newacronym{wmse}{MSE}{weighted mean square error}

\newacronym{ls}{LS}{least squares}
\newacronym{wls}{WLS}{weighted least squares}

\newacronym{mmse}{MMSE}{minimum mean square error}
\newacronym{zf}{ZF}{zero-forcing}
\newacronym{ser}{SER}{symbol error rate}
\newacronym{mf}{MF}{matched filter}

\newacronym{mrt}{MRT}{maximum ratio transmission}

\newacronym{bs}{BS}{base station}
\newacronym{los}{LoS}{line-of-sight}
\newacronym{dac}{DAC}{digital-to-analog converter}
\newacronym{sdp}{SDP}{semidefinite programming problem}

\newacronym{qsdp}{QSDP}{quadratic semidefinite programming problem}
\newacronym{wf}{WF}{Wiener filter}
\newacronym{csi}{CSI}{channel state information}
\newacronym{qam}{QAM}{quadrature amplitude modulation}
\newacronym{doa}{DOA}{direction-of-arrival}
\newacronym{crb}{CRB}{Cramer-Rao Bound}
\newacronym{lfm}{LFM}{linear frequency modulation}
\newacronym{sdr}{SDR}{semi-definite relaxation}

\newacronym{ismr}{ISMR}{integrated sidelobe to mainlobe ratio}

\newacronym{tdd}{TDD}{time-division-duplex}
\newacronym{dft}{DFT}{discrete Fourier transform}
\newacronym{apes}{APES}{amplitude and phase estimation}
\newacronym{sir}{SIR}{signal-to-interference ratio}
\newacronym{is}{IS}{interference suppression}
\newacronym{nmse}{NMSE}{normalized mean-square-error}
\newacronym{mvdr}{MVDR}{minimum variance distortionless response}

\newacronym{altmin}{AltMin}{alternating minimization}

\newacronym{gla}{GLA}{Griffin-Lim algorithm}
\newacronym{kl}{KL}{Kurdyka-\L{}ojasiewicz}

\newacronym{mmpar}{MMPAR}{minimum mainlobe power allocation ratio}
\newacronym{mlf}{MLF}{mainlobe flatness}

\newcommand{\argmin}{\arg\min}

\DeclareMathOperator{\re}{\mathrm{Re}} 
\DeclareMathOperator{\trace}{\mathrm{Tr}} 

\newcommand\figWidth{6} 

\newtheorem{theorem}{Theorem}
\newtheorem*{theorem*}{Theorem}

\newlength\myindent
\newlength\myindentt
\usepackage{tikz}

\newcommand\submittedtext{%
  \footnotesize This work has been submitted to the IEEE for possible publication. Copyright may be transferred without notice, after which this version may no longer be accessible.}

\newcommand\submittednotice{%
\begin{tikzpicture}[remember picture,overlay]
\node[anchor=south,yshift=10pt] at (current page.south) {\fbox{\parbox{\dimexpr0.65\textwidth-\fboxsep-\fboxrule\relax}{\submittedtext}}};
\end{tikzpicture}%
}

\begin{document}

\title{Transmit Beamformer Design for Beampattern Synthesis in Phased-Array Radar Systems}
%
%
%
\author{Berkan Kilic,
        Kenan Turbic,~\IEEEmembership{Member,~IEEE,}
        and
        S\l{}awomir Sta\'nczak,~\IEEEmembership{Senior Member,~IEEE}%
\thanks{%
%
%
This work was partially supported by the Federal Ministry of Research, Technology and Space (BMFTR, Germany) in the “Souverän. Digital. Vernetzt.” programme, joint projects xG-RIC (grant number: 16KIS2429K) and SENSATION (grant number: 16KIS2526).
%
(\textit{Corresponding author: {B. Kilic}.})
}%
    \thanks{
    The authors are with the Wireless Communications and Networks Department, Fraunhofer Heinrich Hertz Institute (HHI), 10587, Berlin, Germany (e-mail: berkan.kilic@hhi.fraunhofer.de, kenan.turbic@hhi.fraunhofer.de, slawomir.stanczak@hhi.fraunhofer.de).
    }%
    \thanks{S. Sta\'nczak is also with Technische Universit{\"a}t Berlin, 10587, Berlin, Germany.
    }%
}

\markboth{}%
{Kilic et al.: Transmit Beamformer Design for Beampattern Synthesis in Phased-Array Radar Systems}
\maketitle

\submittednotice

\IEEEpeerreviewmaketitle

\begin{abstract}
This paper presents a phased-array transmit beamformer design for beampattern matching and interference suppression. We propose an optimization framework jointly addressing these two objectives. The resulting non-convex problem is solved via alternating minimization, where each subproblem admits a closed-form solution, one of which recovers the classical maximum-gain and null-steering phased-array beamformers. To design an algorithm ensuring theoretical convergence guarantees, the beamformer design is then reformulated as a non-convex set feasibility problem, closely related to the original formulation. This problem is addressed using a provably convergent projected gradient descent method and further enhanced with acceleration techniques that substantially improve empirical convergence. Extensive simulations validate the excellent performance and computational efficiency of the proposed method.
\end{abstract}


\glsresetall 

\section{Introduction}
\label{Sec:Intro}

\IEEEPARstart{B}{eampattern} synthesis plays a fundamental role in modern multi-antenna radar and wireless communication systems, providing spatial selectivity that enhances system robustness in the presence of noise and interference \cite{tse2005fundamentals, richards2005fundamentals}. In the context of \gls{tx} beampattern synthesis, the main objective is to efficiently allocate the limited \gls{tx} power budget and focus radiated energy toward spatial directions corresponding to anticipated radar targets, under the free-space propagation assumption \cite{li2008mimo}.

In conventional phased-array architectures, \gls{tx} beamformers are typically realized through analog components. In contrast, \gls{mimo} radars employ fully digital antenna arrays \cite{bjornson2019massive}, enabling simultaneous transmission of independent waveforms from their individual elements. Conventional phased-array radars, by comparison, are restricted to radiating scaled versions of a single waveform \cite{li2008mimo}.

The design flexibility of \gls{mimo} radars has made \gls{tx} beampattern synthesis a central topic in radar and recently emerging \gls{isac} beamforming studies \cite{fuhrmann2008transmit, stoica2007probing, khabbazibasmenj2014efficient, zhang2015mimo, zhang2022min, fan2018constant, hua2013mimo, xu2015colocated, aubry2016mimo, raei2022mimo, guo2024transmit, kilic_jsac, kilic_letter}. 
\Gls{tx} beampatterns are commonly evaluated in terms of sidelobe suppression and mainlobe energy concentration \cite{xu2015colocated, aubry2016mimo}. 
Moreover, it is desirable for the beampattern to exhibit low fluctuation within the mainlobe regions \cite{elliott1985array, gemechu2019beampattern, wang2003optimal}. These metrics are widely employed as optimization objectives for \gls{tx} beampattern synthesis \cite{hua2013mimo, xu2015colocated, aubry2016mimo, fan2018constant, raei2022mimo, guo2024transmit}. 

Beyond these considerations, flexibility in mainlobe beamwidth is essential to accommodate uncertainty in prior information regarding target locations \cite{zhou1999pattern}. In practical radar applications, wider beams are typically preferred in the search mode to cover larger spatial regions, whereas narrower beams are used in the track mode to achieve higher array gain \cite{richards2005fundamentals}.

By considering aforementioned design aspects, a widely adopted strategy for \gls{tx} beampattern synthesis design involves specifying a desired reference \gls{tx} beampattern and minimizing the deviation of the synthesized solution from this target pattern \cite{zhou1999pattern, wang2003optimal}. This methodology, commonly known as \gls{bpm} \cite{stoica2007probing}, provides enhanced modeling flexibility by enabling the designer to explicitly define the desired radiation pattern in accordance with system-level objectives and operational constraints. 

Another advantage of the \gls{bpm} framework in the \gls{mimo} radar setting is that it often admits a convex problem formulation. Unlike phased-array radars, whose waveform correlation matrix must be rank-one, \gls{mimo} radar systems do not impose such non-convex rank constraints, enabling the use of general-purpose convex optimization solvers \cite{diamond2016cvxpy}. As a result, the \gls{bpm} criterion has been extensively employed for \gls{mimo} radar waveform design \cite{fuhrmann2008transmit, stoica2007probing ,khabbazibasmenj2014efficient, zhang2015mimo, fan2018constant, zhang2022min}, as well as in the context of \gls{isac} beamforming \cite{liu2018mu, wu2022mimo, hua2023optimal}. \looseness=-1

The conventional phased-array beamforming strategies aim to maximize array gain in a specific direction and, when required, place nulls at interference directions to improve robustness against interferers \cite{brookner1986adaptive}. To illuminate wide beamwidths or multiple mainlobe regions, distinct beamformers are applied at successive time instances \cite{friedlander2012mimo, duly2013time}. This approach has been widely adopted over decades due to its simple hardware implementation and well-understood adaptation to diverse operational scenarios \cite{daum2009mimo}. However,
these practical benefits come with a critical limitation: the inability to simultaneously illuminate the entire intended spatial region, which is often misconceived as a key advantage of \gls{mimo} radar systems over phased-array radars \cite{friedlander2012mimo}. 

To extend this flexibility to phased-array radar systems, \gls{sdr} techniques can be applied to \gls{mimo} radar formulations \cite{stoica2007probing}. While this suboptimal relaxation yields a convex formulation, it comes at the cost of increased problem dimensionality and necessitates solving potentially large-scale \gls{sdp} instances. Such problems are generally computationally intensive and often become infeasible in high-dimensional setups, such as those encountered in systems deploying massive antenna arrays \cite{kilic_jsac, globecom_bk}. Furthermore, the practical performance of \gls{sdr}-based methods is often unsatisfactory \cite{stoica2007probing}.

An alternative strategy is to formulate the \gls{bpm} objective directly in terms of phased-array beamformer coefficients \cite{tranter2017fast, arora2021efficient, zhang2021fast, vu2023local}. This approach retains the flexibility of the \gls{bpm} approach while avoiding the drawbacks of \gls{sdr}-based methods. However, the rank-one constraint intrinsic to phased-array beamforming renders the problem non-convex, thereby posing significant challenges from an optimization perspective.  

In this work, we focus on phased-array \gls{tx} beamformer design and propose a novel optimization framework that jointly incorporates \gls{bpm} and interference suppression objectives, with interference mitigated via radiation nulling across interference regions. The resulting formulation leads to a non-convex and non-smooth optimization problem, which we address using alternating minimization and split it into two subproblems \cite{beck2017first}. Despite being non-convex, both subproblems admit closed-form optimal solutions. We further show that special cases of the solution to one of these subproblems recover the well-known maximum-gain and null-steering phased-array radar beamformers \cite{friedlander2012mimo}, thereby establishing a direct connection between the proposed \gls{bpm}-based design and conventional phased-array beamformers, a link previously unrecognized in the literature.

Although the subproblems have closed-form solutions, theoretical convergence guarantees for alternating minimization in non-convex settings are limited \cite{attouch2010proximal, hesse2015proximal}. To address this challenge, we reformulate the proposed \gls{tx} beamformer design as a non-convex set feasibility problem \cite{luke2019optimization}. This reformulation enables the development of a solution method based on projected gradient descent that generates a sequence provably converging to a critical point for any initialization.
Furthermore, we introduce an accelerated variant of this method that employs adaptively restarted momentum terms \cite{beck2017first, o2015adaptive}. 

Numerical results demonstrate that, despite having fewer degrees of freedom, the proposed approach achieves lower \gls{psl} and \gls{isl} values, indicating superior \gls{tx} beampattern synthesis performance, while offering significant computational advantages over a widely adopted \gls{mimo} radar design method. We also validate the scalability of the proposed method to large antenna array configurations and demonstrate its effectiveness in an end-to-end radar direction-of-arrival estimation scenario.

The remainder of this manuscript is organized as follows. Section \ref{Sec:SystemModel} presents the system model, and Section~\ref{Sec:DesignOptCriteria} introduces the adopted design optimization criteria for phased-array \gls{tx} beamformer design. The proposed optimization formulation and design methodologies are presented in Section \ref{Sec:ProposedMethod}. Numerical results are provided in Section \ref{Sec:NumericalResults}, followed by concluding remarks in Section \ref{Sec:Conclusions}.

The following notational convention is adopted in the paper.
{$\mathbb{C}^{M\times N}$ and $\mathbb{R}^{M\times N}$ denote the sets of $M \times N$ {complex and real matrices}, respectively. $\mathbb{N}$ denotes the set of natural numbers (positive integers).}
Matrices and vectors are denoted by upper- and lower-case bold letters, respectively.
$x_i$ denotes the $i$-th element of a vector $\boldsymbol{x}$, $\boldsymbol{x}_j$ denotes the $j$-th column of a matrix $\boldsymbol{X}$, while $X_{ij}$ denotes the element in the $i$-th row and $j$-th column of $\boldsymbol{X}$.
$\Vert \boldsymbol{x} \Vert$ denotes the $\ell_2$-norm of $\boldsymbol{x}$. $\Vert \boldsymbol{X} \Vert_F$ and $\Vert \boldsymbol{X} \Vert$ denote the Frobenius norm and the spectral norm of $\boldsymbol{X}$, respectively. Transpose, conjugate and conjugate-transpose (Hermitian) of $\boldsymbol{X}$ are denoted by $\boldsymbol{X}^T$, $\boldsymbol{X}^*$ and $\boldsymbol{X}^H$, respectively, while its trace is denoted by $\mathrm{Tr}(\boldsymbol{X})$. 
$\boldsymbol{X} \succeq \boldsymbol{0}$ indicates positive semi-definiteness of a (Hermitian) matrix $\boldsymbol{X}$.  
$\boldsymbol{I}_N$ is the identity matrix of size $N$. $|c|$ is the modulus of $c\in \mathbb{C}$. The cardinality of a set $\Phi$ is denoted by $|\Phi|$. 
$\mathbb{E}[\,.\,]$ denotes statistical expectation. 


\section{System Model}
\label{Sec:SystemModel}
We consider a narrowband monostatic radar system with a \gls{tx} array of $M_T$ antennas and a \gls{rx} array of $M_R$ antennas, both operating in the far-field region. To simplify the analysis, we assume that the signal propagation is confined to the horizontal plane. The array steering vectors characterizing \gls{tx} and \gls{rx} array responses are denoted by $\boldsymbol{a}_T(\phi) \in \mathbb{C}^{M_T \times 1}$ and $\boldsymbol{a}_R(\phi) \in \mathbb{C}^{M_R \times 1}$, respectively, where $\phi$ denotes the azimuth direction relative to the antenna array boresight ($\phi = 0^\circ$). 

Let $\boldsymbol{x}(t) \in \mathbb{C}^{M_T \times 1}$ denote the \gls{tx} signal at time $t$. The baseband representation of \gls{rx} signal reflected from a point target at azimuth direction $\phi$ can be then written as \cite{li2008mimo} \looseness=-1
\begin{equation}
\label{Eqn:RadarRx}
\boldsymbol{y}_R(t) \coloneqq q(\phi)\boldsymbol{a}^*_R(\phi)\boldsymbol{a}_T^H(\phi)\boldsymbol{x}(t) + \boldsymbol{n}_R(t),
\end{equation}
where $q(\phi) \in \mathbb{C}$ denotes the target reflection coefficient, and $\boldsymbol{n}_R(t) \sim \mathcal{CN}(\boldsymbol{0}, \sigma_n^2\boldsymbol{I}_{M_R})$ consists of i.i.d. zero-mean additive complex Gaussian noise samples.

The \gls{tx} beampattern, characterizing the directional radiation intensity and thus affecting the \gls{rx} \gls{snr}, is given by \cite{friedlander2012mimo}
\begin{equation}
    \label{Eqn:TransmitBeampattern}
    P(\phi) \coloneqq \boldsymbol{a}_T^H(\phi) \boldsymbol{R}_T \boldsymbol{a}_T(\phi), 
\end{equation}
where
\begin{equation}
    \label{Eqn:MIMOWaveformCorr}
    \boldsymbol{R}_T  \coloneqq \frac{1}{T}\int_{0}^{T} \boldsymbol{x}(t)\boldsymbol{x}^H(t)\, dt,
\end{equation}
denotes the \gls{tx} signal correlation matrix and $T$ is the coherent processing interval. The average \gls{tx} power of the radar system is computed as
\begin{equation}
    \mathrm{Tr}(\boldsymbol{R}_T) = \frac{1}{T}\int_{0}^{T} \Vert \boldsymbol{x}(t)\Vert^2 dt.
\end{equation}

In the case of a phased-array radar, the \gls{tx} array is limited to emitting scaled versions of a single waveform $s(t)$ from its antennas. Then, we can write the \gls{tx} signal as 
\begin{equation}
    \label{Eqn:PhasedWaveform}
    \boldsymbol{x}(t) = \boldsymbol{w}_T{s}(t),
\end{equation}
where $\boldsymbol{w}_T \in \mathbb{C}^{M_T \times 1}$ is the \gls{tx} beamformer vector.
Without loss of generality, we assume that $s(t)$ has unit average power, i.e., 
\begin{equation}
    \label{Eqn:WaveformUnitEnergy}
    \frac{1}{T}\int_{0}^{T} {s}(t){s}^*(t)  dt = \frac{1}{T}\int_{0}^{T} |{s}(t)|^2  dt = 1.
\end{equation}
Then, substituting \eqref{Eqn:PhasedWaveform} into \eqref{Eqn:MIMOWaveformCorr} yields
\begin{equation}
    \label{Eqn:PhasedArrayCorr}
    \boldsymbol{R}_T = \boldsymbol{w}_T \boldsymbol{w}_T^{H}, 
\end{equation} 
where $\mathrm{rank}(\boldsymbol{R}_T) = 1$.

Unlike phased array radars, \gls{mimo} radars can transmit independent waveforms from their individual antenna elements.
Consequently, in the \gls{mimo} radar case, $\boldsymbol{R}_T$ can have rank up to $M_T$, whereas in a phased-array radar system, it is intrinsically rank-one, as observed from \eqref{Eqn:PhasedArrayCorr}.

In this study, our focus is on \gls{tx} beamforming design for phased-array radar systems, i.e., the design of $\boldsymbol{w}_T$. 
Our proposed methodology is decoupled from the choice of $s(t)$, whose design is based on the application scenario \cite{sun2014analysis} (e.g., desired range/Doppler sidelobe levels), where the conventional choice for radar systems is the chirp signal \cite{richards2005fundamentals}. 

\begin{figure}[t]
    \centering
    \includegraphics[width=0.95\columnwidth]{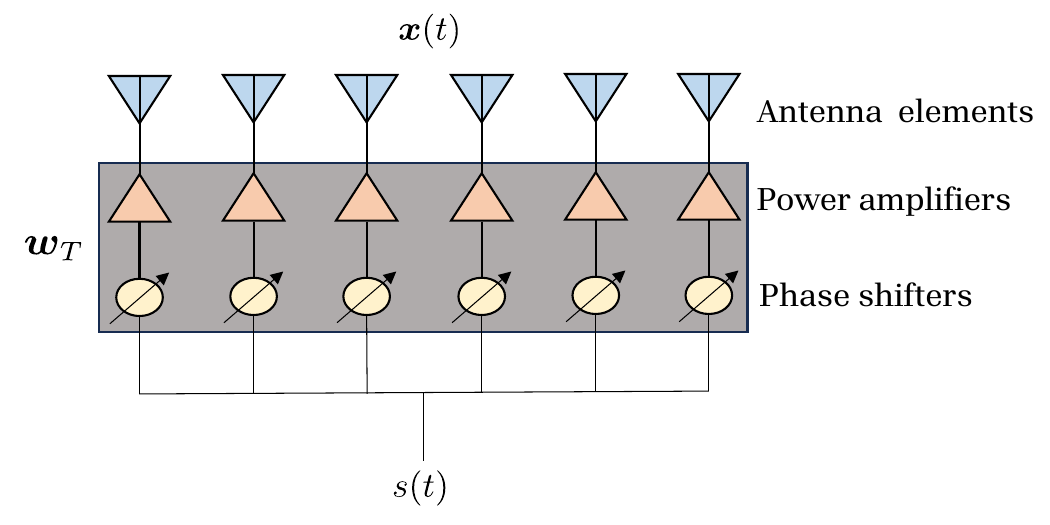}
    \caption{Phased-array architecture for \gls{tx} beamforming.}
    \label{beamformer_tp}
\end{figure}

A simplified phased-array architecture for the hardware implementation of $\boldsymbol{w}_T$ is shown in Fig.~\ref{beamformer_tp}. In this setup, each antenna is connected to a dedicated power amplifier and a phase shifter, which are assumed to enable independent control of both the amplitudes and phase shifts of the coefficients of $\boldsymbol{w}_T$. \looseness=-1

\section{Design Optimization Criteria}
\label{Sec:DesignOptCriteria}
As discussed in Section \ref{Sec:Intro}, a commonly adopted approach for beampattern synthesis in radar systems is based on the \gls{bpm} criterion. In addition, the radar system is required to be resilient to jammer and clutter effects, necessitating effective interference suppression capabilities. 
Here we present the optimization frameworks to meet these objectives.

\subsection{Beampattern Matching Objective}
\label{Sec:DesignCriteriaBPM}
The primary goal in \gls{bpm} is to design the \gls{tx} correlation matrix $\boldsymbol{R}_T$ such that the resulting \gls{tx} beampattern $P(\phi)$ closely approximates a desired pattern. To formulate this objective, let $\Phi_g = \{ \bar{\phi}_i \}_{i=1}^{N_{\phi}}$ be the set denoting the angular grid covering the field of view, with $N_{\phi}$ being the grid size. The mainlobe region set $\Phi_m \subseteq \Phi_g$ specifies the directions where targets are expected, which is assumed to be non-empty. The complementary set $\Phi_s$ denotes the sidelobe region set, where $\Phi_m \cup \Phi_s = \Phi_g$ and $\Phi_m \cap \Phi_s = \varnothing$. Assuming equal priority across all directions in $\Phi_m$, the normalized desired \gls{tx} beampattern can be defined as \cite{fuhrmann2008transmit, stoica2007probing} 
\begin{equation}
    \label{Eqn:DesiredBeampatternDefn}
    P_d(\bar{\phi}_i) =
    \begin{cases}
        1,~ &\text{if } \bar{\phi}_i \in \Phi_m, \\
        0,~ &\text{otherwise},
    \end{cases}
\end{equation}
which means, all available energy is desired to be fully focused toward mainlobe regions.

The actual \gls{tx} beampattern $P(\bar{\phi}_i)$ should approximate $\eta P_d(\bar{\phi}_i)$ well, over all $\bar{\phi}_i \in \Phi_g$, with $ {\eta} \ge 0$ introduced to match the scalings of $P_d(\bar{\phi}_i)$ and $P(\bar{\phi}_i)$. This leads to the commonly used \gls{bpm} objective \cite{stoica2007probing, he2012waveform, liu2018mu, wu2022mimo, wang2024robust, li2007beampattern}, defined as: 
\begin{align}
    \label{Eqn:Jm_bpm}
    J^{m}_{\mathrm{bpm}}(\boldsymbol{R}_T, \eta)& \coloneqq \sum_{i=1}^{N_{\phi}} \left( P(\bar{\phi}_i) -  {\eta} P_d(\bar{\phi}_i) \right)^2 \nonumber \\
    = &\sum_{i=1}^{N_{\phi}} \left( \boldsymbol{a}_T^H(\bar{\phi}_i)\boldsymbol{R}_T\boldsymbol{a}_T(\bar{\phi}_i) -  {\eta} P_d(\bar{\phi}_i) \right)^2.
\end{align}

For practical feasibility, $\boldsymbol{R}_T$ must also satisfy additional conditions. Let $P_T > 0$ denote the total power budget of the system. 
Assuming the full utilization of the available energy, we have $\mathrm{Tr}(\boldsymbol{R}_T)=P_T$. Moreover, since it is a correlation matrix, $\boldsymbol{R}_T$ must be positive semidefinite by definition. Then, the following optimization problem is typically formulated for \gls{mimo} radar \gls{tx} beamforming design \cite{li2008mimo}:
\begin{equation}
    \label{Eqn:SDP_Formulation}
    \min_{\boldsymbol{R}_T, \eta \ge 0} J^{m}_{\mathrm{bpm}}(\boldsymbol{R}_T, \eta)
    \quad \text{s.t.} \quad \trace(\boldsymbol{R}_T) = P_T,\;
    \boldsymbol{R}_T\succeq \boldsymbol{0}.  
\end{equation}
While \eqref{Eqn:SDP_Formulation} is convex, phased-array radar systems impose the additional rank-one constraint on the waveform correlation matrix, which renders the resulting problem non-convex. 

This non-convex rank constraint is typically addressed via \gls{sdr} \cite{luo2010semidefinite}, i.e., the rank-one approximation of the solution of \eqref{Eqn:SDP_Formulation} is used to construct the phased-array beamformer. 
This approach lifts the original beamformer design from an $M_T$-dimensional vector $\boldsymbol{w}_T$ to an optimization over the $M_T \times M_T$ matrix $\boldsymbol{R}_T$, leading to a quadratic increase in problem dimension. As a result, \gls{sdr}-based methods incur high computational complexity and often exhibit poor beampattern synthesis performance~\cite{stoica2007probing}.

\begin{figure}[t]
    \centering
    \includegraphics[width=6 cm]{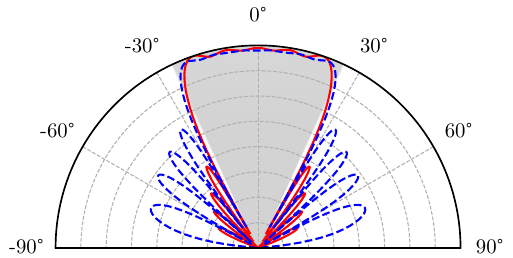}
    \caption{Desired (gray-shaded) and synthesized beampatterns by two different beamformers, shown with red and blue lines.}
    \label{polar_bp}
\end{figure}

A more direct approach is to directly define the \gls{bpm} cost on $\boldsymbol{w}_T$. To this end, by the column-wise concatenation of steering vectors corresponding to each grid point, we define the array manifold matrix: 
\begin{align}
    \label{Eq:D}
    \boldsymbol{D} \coloneqq \left[\boldsymbol{a}_T(\bar{\phi}_1)~ \boldsymbol{a}_T(\bar{\phi}_2)~...~\boldsymbol{a}_T(\bar{\phi}_{N_{\phi}})\right], 
\end{align}
which is assumed to be full-rank. 
By also defining $b_i \coloneqq   [{P_d(\Bar{\phi}_i)}]^{1/2}$ for all $1 \le i \le N_{\phi}$ and writing $\boldsymbol{R}_T = \boldsymbol{w}_T\boldsymbol{w}_T^H$, we ideally obtain $J^{m}_{\mathrm{bpm}}(\boldsymbol{R}_T, \eta)=0$ in \eqref{Eqn:Jm_bpm} when the following holds 
\begin{equation}
     \label{Eqn:PR_Equality}
     |\boldsymbol{D}^H(\Bar{\eta}^{-1}\boldsymbol{w}_T) | = \boldsymbol{b},
\end{equation}
where $ \Bar{\eta} \coloneqq \sqrt{\eta}$ and $|.|$ is the entry-wise magnitude operator. This then motivates us to define the \gls{bpm} cost in the phased-array radar case as 
\begin{equation}
    \label{Eqn:Jp_bpm}
    J^{p}_{\mathrm{bpm}}(\boldsymbol{w}_T, \Bar{\eta})\coloneqq  \Vert |\boldsymbol{D}^H(\Bar{\eta}^{-1}\boldsymbol{w}_T) | - \boldsymbol{b} \Vert^2, 
\end{equation}
which, in related forms, has been adopted in \cite{tranter2017fast, arora2021efficient}.

Fig.~\ref{polar_bp} illustrates the generated and desired beampatterns for an example scenario, comparing two beamformer design algorithms. From the figure, it is observed that one of the generated beampatterns has significantly lower sidelobes, highlighting the critical impact of \gls{tx} beamformer design.

\subsection{Interference Suppression Objective}
In practical radar environments, both intentional and unintentional interference can severely degrade target detection and parameter estimation performance at the radar \gls{rx} \cite{richards2005fundamentals}.
In addition to the interference from active sources, strong reflections from nearby objects (clutter) can also act as passive interferers, masking weaker echos from more distant targets.
The \gls{tx}-side interference suppression should minimize illumination of such passive interferers, while also avoiding unnecessary radiation toward directions of active interferers.

Motivated by these considerations, it is desired for the synthesized \gls{tx} beampattern to exhibit deep notches along the directions of interfering sources (e.g., jammers)~\cite{brookner1986adaptive}. 
To formulate this objective, let $\Phi_J \subseteq \Phi_g$ denote the subset of (known) interference directions, which can include potentially several but separated single points (spot jamming) or multiple angular regions (barrage jamming) \cite{stoica2011optimization}. 
The corresponding objective is nulling $P(\Bar{\phi}_i)$ for all $\Bar{\phi_i} \in \Phi_J$. 

Then, we define the following cost function for the objective of interference suppression: 
\begin{equation}
    \label{Eqn:Jp_is}
    J^{p}_{\mathrm{is}}(\boldsymbol{w}_T, \Bar{\eta}) 
    \coloneqq  \sum_{\Bar{\phi}_i \in \Phi_J} |\boldsymbol{a}_T^H(\Bar{\phi}_i) (\Bar{\eta}^{-1}\boldsymbol{w}_T)|^2, 
\end{equation}
resulting in the following overall phased-array radar objective jointly incorporating \gls{bpm} and interference suppression terms:  
\begin{equation}
    \label{Eqn:Jp_combined}
    J^{p}(\boldsymbol{w}_T, \Bar{\eta}) \coloneqq  J^{p}_{\mathrm{bpm}}(\boldsymbol{w}_T, \Bar{\eta}) + \Bar{\omega} J^{p}_{\mathrm{is}}(\boldsymbol{w}_T, \Bar{\eta}),
\end{equation}
where $\Bar{\omega} \ge 0$ is the weight assigned to interference suppression objective that should be selected based on the power of the interfering source, i.e., deeper notches are formed as $\Bar{\omega}$ gets larger.

The interference directions are typically assumed to intersect with sidelobe regions \cite{brookner2017mimo}, i.e., $\Phi_J \subseteq\Phi_s$, which we also assume here. Then, we can equivalently rewrite \eqref{Eqn:Jp_combined} as 
\begin{equation}
    \label{Eqn:FinalCostBPM_IS}
    J^{p}(\boldsymbol{w}_T, \Bar{\eta}) = \Vert |\boldsymbol{\Bar{D}}^H(\Bar{\eta}^{-1}\boldsymbol{w}_T) | - \boldsymbol{b} \Vert^2,
\end{equation}
where the columns of $\boldsymbol{\Bar{D}}$ are defined as 
\begin{equation}
    \label{Eqn:DbarDefn}
    \boldsymbol{\Bar{d}}_i \coloneqq \begin{cases}
        {\omega}\boldsymbol{a}_T(\Bar{\phi}_i),~ &\text{if } \Bar{\phi}_i \in \Phi_J, \\
        \boldsymbol{a}_T(\Bar{\phi}_i),~ &\text{otherwise}, 
    \end{cases}
\end{equation}
with $\omega \coloneqq \sqrt{\Bar{\omega} + 1}$. 

\section{Proposed Methodology}
\label{Sec:ProposedMethod}
In this section, we present our proposed methodology for phased-array \gls{tx} beamformer design. In Section \ref{Sec:ClassicalPhasedArray}, we first revisit conventional maximum-gain and null-steering phased-array beamformers. Then, in Section \ref{Sec:ProposedOptFramework}, we present the proposed optimization framework that incorporates \gls{bpm} and interference suppression objectives under a total power constraint. 
To solve this non-convex and non-smooth problem, we introduce an alternating minimization approach in Section~\ref{Sec:SolnAltMin}, which splits the problem into two subproblems that admit closed-form solutions, one of which recovers the aforementioned classical beamformers as a special case.
Building on the alternating minimization solution, we reformulate the problem as a non-convex set feasibility problem in Section~\ref{Sec:GriffinLim}. This formulation enables the development of a provably convergent\footnote{Here, we refer to the convergence of the whole sequence generated by the algorithm to a critical point of the tackled problem.} projected gradient method in Section~\ref{Sec:ProjGrad}, followed by its accelerated variant in Section~\ref{Sec:AccProjGrad}. Finally, Section~\ref{Sec:ImplementationAspects} presents computational complexity analysis and practical implementation details.

\subsection{Conventional Phased-Array Beamformers}
\label{Sec:ClassicalPhasedArray}
When the design objective is to maximize the \gls{tx} beampattern gain toward a desired direction under a total-power constraint, the optimal beamformer is the conventional maximum-gain phased-array beamformer\footnote{This is the \gls{tx} counterpart of the conventional Bartlett \gls{rx} beamformer.} \cite{richards2005fundamentals}, given by
\begin{equation}
    \label{Eqn:PhasedBF_maxSNR}
    \boldsymbol{w}_T^\diamond = \sqrt{P_T}\,
    \frac{\boldsymbol{a}_T(\bar{\phi})}{\|\boldsymbol{a}_T(\bar{\phi})\|},
\end{equation}
where \(\bar{\phi}\) denotes the target direction. In this case, we are concerned with maximizing the gain in one direction, regardless of the sidelobes. 

In the presence of interferers, a conventional approach is to steer the mainbeam toward the desired direction while placing nulls toward the interference directions. Let \(\boldsymbol{D}_J\) be formed by column-wise concatenation of the corresponding steering vectors \(\boldsymbol{a}_T(\bar{\phi}_i)\), \(\bar{\phi}_i \in \Phi_J\). We assume that the columns of $\boldsymbol{D}_J$ are linearly independent and that 
$\boldsymbol{a}_T(\Bar{\phi})$ is linearly independent of the columns of $\boldsymbol{D}_J$. Then the corresponding null-steering phased-array beamformer is given by \cite{friedlander2012mimo}
\begin{equation}
    \label{Eqn:PhasedBF_nullSteer}
    \boldsymbol{w}_T^\diamond =
    \sqrt{P_T}\,
    \frac{\boldsymbol{\Pi}_J \boldsymbol{a}_T(\bar{\phi})}
    {\|\boldsymbol{\Pi}_J \boldsymbol{a}_T(\bar{\phi})\|},
\end{equation}
where
\begin{equation}
    \boldsymbol{\Pi}_J \coloneqq
    \boldsymbol{I}_{M_T}
    - \boldsymbol{D}_J
    (\boldsymbol{D}_J^H \boldsymbol{D}_J)^{-1}
    \boldsymbol{D}_J^H.
\end{equation}
The beamformer in \eqref{Eqn:PhasedBF_nullSteer} is the well-known null-steering phased-array beamformer. Note that without considering the interference effects, \eqref{Eqn:PhasedBF_nullSteer} reduces to \eqref{Eqn:PhasedBF_maxSNR}. 

\subsection{Proposed Optimization Framework}
\label{Sec:ProposedOptFramework}
To jointly address beampattern matching and interference suppression, we adopt the objective function in \eqref{Eqn:FinalCostBPM_IS}. Assuming the full utilization of the available energy, we write: 
\begin{equation}
    \mathrm{Tr}(\boldsymbol{R}_T) = \Vert \boldsymbol{w}_T \Vert^2 = P_T,
\end{equation}
and then formulate the following unified optimization problem for phased-array radar \gls{tx} beamformer design:
\begin{equation}
    \label{Eqn:Phased_Formulation_proposed}
    \min_{\boldsymbol{w}_T, \Bar{\eta} \ge 0} \Vert |\boldsymbol{\Bar{D}}^H(\Bar{\eta}^{-1}\boldsymbol{w}_T) | - \boldsymbol{b} \Vert^2
    \quad \text{s.t.} \quad \Vert \boldsymbol{w}_T \Vert^2 = P_T.    
\end{equation}
Unlike the convex \gls{mimo} radar beamformer design in \eqref{Eqn:SDP_Formulation}, \eqref{Eqn:Phased_Formulation_proposed} is both non-convex and non-smooth, since the entry-wise magnitude operator $|.|$ in the objective introduces non-smoothness and non-convexity, while the total-power constraint $\Vert \boldsymbol{w}_T \Vert^2 = P_T$ is also non-convex. In the following sections, we focus on developing computationally efficient methods with provable convergence to solve \eqref{Eqn:Phased_Formulation_proposed}.

\subsection{Solution Method Based on Alternating Minimization}
\label{Sec:SolnAltMin}
Without the power constraint and the additional optimization variable $\Bar{\eta}$, \eqref{Eqn:Phased_Formulation_proposed} resembles the well-known phase retrieval problem \cite{fienup1982phase}, for which alternating minimization is a widely used solution method \cite{netrapalli2013phase}. To apply alternating minimization method to solve \eqref{Eqn:Phased_Formulation_proposed}, we note that its objective function becomes zero when the following equality holds 
\begin{equation}
    \label{Eqn:PR_Equality2}
     \boldsymbol{\Bar{D}}^H(\Bar{\eta}^{-1}\boldsymbol{w}_T)  = \boldsymbol{{p}} \odot \boldsymbol{b},
\end{equation}
where $\odot$ denotes the Hadamard product and ${\boldsymbol{{p} }\in\mathbb{C}^{N_{\phi} \times 1}}$ consists of the element-wise phases of $\boldsymbol{\Bar{D}}^H \boldsymbol{w}_T$, i.e., ${\boldsymbol{{p}} = \mathrm{sgn}(\boldsymbol{\Bar{D}}^H\boldsymbol{w}_T)}$, with the $\mathrm{sgn}$ function defined for ${1 \le i \le N_{\phi}}$ as \cite{bauschke2002phase}:
\begin{equation}
    \label{Eqn:sgnDefn}
    [\mathrm{sgn}(\boldsymbol{u})]_i \coloneqq 
        \begin{cases}
         u_i/|u_i|,~&\mathrm{if}~ u_i \ne 0,\\
        1,~ &\mathrm{otherwise}. 
    \end{cases}
\end{equation}
To get rid of $|.|$ operator in \eqref{Eqn:Phased_Formulation_proposed},
by introducing $\boldsymbol{{p}}$ as an additional optimization variable, we cast the following optimization problem
\begin{align}
    \label{Eqn:PR_Formulation}
    &\min_{\boldsymbol{w}_T, \Bar{\eta}, \boldsymbol{p}}\left\Vert \boldsymbol{\Bar{D}}^H(\Bar{\eta}^{-1}\boldsymbol{w}_T) - \boldsymbol{{p}} \odot \boldsymbol{b} \right\Vert^2 \nonumber \\
    &\mathrm{~s.t.~} \Vert \boldsymbol{w}_T \Vert^2 = P_T,~|\boldsymbol{p}|=\boldsymbol{1},~\Bar{\eta} \ge 0,
\end{align}
where $\boldsymbol{1}$ denotes the vector of all ones. 

Now we are in a position to introduce the alternating minimization procedure, with $k\ge 1$ denoting the iteration number. Given $\boldsymbol{w}_T^{k-1}$ and $\Bar{\eta}^{k-1}$, the minimizer of \eqref{Eqn:PR_Formulation} in the $k$-th iteration is obtained as \cite{netrapalli2013phase}
\begin{align}
    \label{Eqn:UpdateP}
    \boldsymbol{p}^{k} &= \mathrm{sgn}(\boldsymbol{\Bar{D}}^H\boldsymbol{w}_T^{k-1}) \nonumber \\ 
    &\in \argmin_{|\boldsymbol{p}|=\boldsymbol{1}} \left \Vert (\Bar{\eta}^{k-1})^{-1}\boldsymbol{\Bar{D}}^H\boldsymbol{w}_T^{k-1} - \boldsymbol{p}\odot \boldsymbol{b} \right\Vert^2 . 
\end{align}
Given $\boldsymbol{p}^{k}$, we perform the minimization jointly over $\boldsymbol{w}_T$ and $\eta$: \looseness=-1
\begin{equation}
    \label{Eqn:OptOverx}
    \left\{ \boldsymbol{w}_T^{k}, \Bar{\eta}^{k} \right\} \in \argmin_{\substack{\Vert \boldsymbol{w}_T \Vert^2 = P_T, \\ \Bar{\eta} \ge 0}} 
\left\Vert \boldsymbol{\Bar{D}}^H \Bar{\eta}^{-1}\boldsymbol{w}_T 
- \boldsymbol{p}^{k} \odot \boldsymbol{b} \right\Vert^2.  
\end{equation}
We obtain the closed-form solution to \eqref{Eqn:OptOverx} as \cite[Thm. 1]{zhang2018multibeam}
\begin{equation}
    \label{Eqn:Eta_x_update}
    \boldsymbol{w}_T^{k} = \Bar{\eta}^{k} (\boldsymbol{\Bar{D}}^H)^{\dagger}\left(\boldsymbol{p}^k \odot \boldsymbol{b}\right),
\end{equation}
where
$(\boldsymbol{\Bar{D}}^H)^\dagger$ is the pseudo-inverse of $\boldsymbol{\Bar{D}}^H$ and
\begin{align}
    \label{Eqn:Eta_eta_update}
    \Bar{\eta}^{k} = \frac{\sqrt{P_T}}{\Vert (\boldsymbol{\Bar{D}}^H)^{\dagger} (\boldsymbol{p}^k \odot \boldsymbol{b}) \Vert}. 
\end{align}
In \eqref{Eqn:Eta_x_update}, $\Bar{\eta}^{k}$ serves as a normalization factor that enforces the total \gls{tx} power constraint.

As demonstrated in Appendix~\ref{Appendix:SpecialCases}, for fixed $\boldsymbol{p}=\boldsymbol{1}$, the solution in~\eqref{Eqn:Eta_x_update} reduces to well-known beamforming strategies for appropriate choices of $\Phi_g$. In particular, choosing  ${\Phi_g = \{\bar{\phi}\}}$ yields the maximum-gain beamformer, while setting ${\Phi_g = \Phi_J \cup \{\bar{\phi}\}}$ recovers the null-steering beamformer, given in~\eqref{Eqn:PhasedBF_maxSNR} and~\eqref{Eqn:PhasedBF_nullSteer}, respectively.

We note that the minimizer given in \eqref{Eqn:Eta_x_update} is valid for both $N_{\phi} \ge M_T$ (overdetermined case) and $N_{\phi} < M_T$ (underdetermined case) with the corresponding pseudoinverse definitions \cite{horn2012matrix}. However, it is typical to assume $N_{\phi} \ge M_T$, i.e., the selected number of grid points covering the field of view is larger than the number of antennas. In the rest of this manuscript, we adopt this assumption. 

\begin{algorithm}[t]
\caption{Alternating projections method to solve 
\eqref{Eqn:OptimizationMultiSet1}}
\begin{algorithmic}[1] 
\label{Algo:GLA}
    \renewcommand{\algorithmicrequire}{\textbf{Input:}}
    \renewcommand{\algorithmicensure}{\textbf{Output:}}
    \REQUIRE $\boldsymbol{A}, \boldsymbol{b}, P_T$
\ENSURE $\boldsymbol{w}_T^\star$ 
    \STATE {Initialize} $\boldsymbol{{z}}^0 \in \mathbb{C}^{N_{\phi} \times 1}$
    \FOR{$k=1,2,...$}
    \STATE Update $\boldsymbol{z}^k$ according to \eqref{Eqn:AltProj}
    \ENDFOR
    \STATE Return $\boldsymbol{w}_T^\star = \mathrm{nrm}\left( \boldsymbol{A}^\dagger \boldsymbol{z}^k \right)$
\end{algorithmic} 
\end{algorithm}
\subsection{Solution Method Based on Alternating Projections}
\label{Sec:GriffinLim}
Although alternating minimization often performs well in practice, its convergence guarantees in non-convex settings remain limited \cite{hesse2015proximal}.
In this section, we present an alternative formulation of \eqref{Eqn:Phased_Formulation_proposed} that is better suited for non-convex and non-smooth optimization.
For the simplicity of notation, we define $\boldsymbol{A} \coloneqq \boldsymbol{\Bar{D}}^H$ with the pseudoinverse 
$\boldsymbol{A}^\dagger = (\boldsymbol{A}^H \boldsymbol{A})^{-1} \boldsymbol{A}^H$
such that $\boldsymbol{A}^\dagger \boldsymbol{A} = \boldsymbol{I}_{M_T}$. 

We first note that the alternating iterations given in \eqref{Eqn:UpdateP}, \eqref{Eqn:Eta_x_update} and \eqref{Eqn:Eta_eta_update} can equivalently be expressed as the following fixed-point iteration: 
\begin{equation}
    \label{Eqn:FixedPoint}
    \boldsymbol{w}_T^{k}=\mathrm{nrm}\left(  \boldsymbol{A}^\dagger (\mathrm{sgn}(\boldsymbol{A}\boldsymbol{w}_T^{k-1}) \odot \boldsymbol{b}) \right),
\end{equation}
where $\mathrm{nrm}$ is the power normalization function, i.e., $\mathrm{nrm}(\boldsymbol{u}) \coloneqq \sqrt{P_T} \boldsymbol{u}/\Vert \boldsymbol{u} \Vert$.

Since the output of the $\mathrm{sgn}$ function remains invariant under scaling with a positive number, the normalization of $\boldsymbol{w}_T^{k-1}$ does not affect the computation of the argument in the $\mathrm{nrm}$ function in \eqref{Eqn:FixedPoint}. 
Therefore, it is not needed to perform the normalization in each step. We can then replace the iterates given in \eqref{Eqn:FixedPoint} with:  
\begin{equation}
    \label{Eqn:FixedPoint2}
    \boldsymbol{\Tilde{w}}_T^{k}=  \boldsymbol{A}^\dagger (\mathrm{sgn}(\boldsymbol{A}\boldsymbol{\Tilde{w}}_T^{k-1}) \odot \boldsymbol{b}),
\end{equation}
and equivalently obtain the normalized solution as ${\boldsymbol{{w}}_T^{q} = \mathrm{nrm}(\boldsymbol{\Tilde{w}}_T^{q})}$, where $q$ denotes the iteration index at which the procedures in \eqref{Eqn:FixedPoint} and \eqref{Eqn:FixedPoint2} are terminated. This observation suggests formulating the optimization directly in terms of $\boldsymbol{\Tilde{w}}_T$, which represents the unnormalized beamformer vector $\boldsymbol{w}_T$. The power normalization is then applied only at the final iteration, providing the basis for the problem formulation we introduce in the following.

To proceed, we first define the following sets \cite{nenov2023accelerated} 
\begin{align}
    \label{Eqn:SetQ}
    &Q \coloneqq \left\{ \boldsymbol{z}\in \mathbb{C}^{N_{\phi} \times 1 } \big|~ |\boldsymbol{z}| = \boldsymbol{b} \right\},  \\
    \label{Eqn:SetC}
    &C \coloneqq \left\{ \boldsymbol{z}\in \mathbb{C}^{N_{\phi} \times 1 } \big|~  \exists\boldsymbol{\Tilde{w}}_T \in\mathbb{C}^{M_T \times 1 }: \boldsymbol{z} = \boldsymbol{A}\boldsymbol{\Tilde{w}}_T \right\}, 
\end{align}
and cast the following non-convex set feasibility problem \cite{luke2019optimization}: 
\begin{equation}
    \label{Eqn:OptimizationMultiSet0}
    \mathrm{Find~} \boldsymbol{{z}} \in C \cap Q.    
\end{equation}
If $\boldsymbol{z} \in C \cap Q$ exists, then there exists $\boldsymbol{\Tilde{w}}_T$ such that
\begin{equation}
    |\boldsymbol{A}\boldsymbol{\Tilde{w}}_T| = \boldsymbol{b},  
\end{equation}
which corresponds to the ideal \gls{bpm} condition in \eqref{Eqn:PR_Equality}. In general, however, $C \cap Q$ may be empty, i.e., \eqref{Eqn:OptimizationMultiSet0} may be infeasible, which we address in Section \ref{Sec:ProjGrad}. 

We can rewrite \eqref{Eqn:OptimizationMultiSet0} in the following unconstrained form:  
\begin{equation}
    \label{Eqn:OptimizationMultiSet1}
    \min_{\boldsymbol{z}} \{ \delta_{Q}(\boldsymbol{z}) +  \delta_{C}(\boldsymbol{z}) \},
\end{equation}
where $\delta_{\mathcal{X}}$ denotes the indicator function of a set $\mathcal{X}$, i.e., 
\begin{equation}
    \delta_{\mathcal{X}}(\boldsymbol{u}) \coloneqq 
    \begin{cases}
         0,~&\mathrm{if}~ \boldsymbol{u} \in \mathcal{X},\\
        +\infty,~ &\mathrm{otherwise}. 
    \end{cases}
\end{equation}
The problems \eqref{Eqn:Phased_Formulation_proposed} and \eqref{Eqn:OptimizationMultiSet1} are closely related, as we clarify in the following. 

A conventional method to solve \eqref{Eqn:OptimizationMultiSet1} is based on alternating projections, which is performed by projecting $\boldsymbol{z}$ onto the sets $Q$ and $C$ in a cyclic manner, where the projector $P_{\mathcal{X}}$ is defined as  
\begin{equation}
    P_{\mathcal{X}}(\boldsymbol{u}) \coloneqq \argmin_{\boldsymbol{v}\in \mathcal{X}} \Vert \boldsymbol{u} - \boldsymbol{v} \Vert,  
\end{equation}
which is a set-valued mapping in general, e.g., for the non-convex set $Q$ defined in \eqref{Eqn:SetQ}. Then, the iterates of the alternating projections method are expressed as 
\begin{equation}
    \label{Eqn:AlternatingProjections}
    \boldsymbol{z}^k \in P_C\left (P_Q(\boldsymbol{z}^{k-1})\right),  
\end{equation}
where $P_Q$ and $P_C$ can be computed as  
\cite{bauschke2002phase, nenov2023accelerated}
\begin{align}
    \label{Eqn:PqSgnRelationship}
    & P_Q(\boldsymbol{z}^{k-1}) \ni \mathrm{sgn}(\boldsymbol{z}^{k-1}) \odot \boldsymbol{b} , \\ &P_C(\boldsymbol{u}) = \boldsymbol{A}\boldsymbol{A}^\dagger\boldsymbol{{u}} .
\end{align}
Then, we express \eqref{Eqn:AlternatingProjections} as the following fixed-point iteration:
\begin{equation}
    \label{Eqn:AltProj}
    \boldsymbol{z}^k = \boldsymbol{A}\boldsymbol{A}^\dagger \left( \mathrm{sgn}(\boldsymbol{z}^{k-1}) \odot \boldsymbol{b} \right) \in P_C\left (P_Q(\boldsymbol{z}^{k-1})\right). 
\end{equation}
Once $\boldsymbol{z}^\star$ is obtained, the solution follows from $\boldsymbol{\Tilde{w}}_T^\star = \boldsymbol{A}^\dagger \boldsymbol{z}^\star$, and is found as $\boldsymbol{w}_T^\star = \mathrm{nrm}(\boldsymbol{\Tilde{w}}_T^\star)$. 
Since the whole sequence $\{ \boldsymbol{z}^k \}_{k \in \mathbb{N}}$ generated by \eqref{Eqn:AltProj} resides in the column space of $\boldsymbol{A}$, there always exists a $\boldsymbol{\Tilde{w}}_T^k$ satisfying $\boldsymbol{z}^k = \boldsymbol{A} \boldsymbol{\Tilde{w}}_T^k$ for all $k \ge 1$, which can be substituted into \eqref{Eqn:FixedPoint2} to obtain \eqref{Eqn:AltProj}. 
Therefore, the iterates generated by the alternating minimization method for solving \eqref{Eqn:Phased_Formulation_proposed} coincide with those produced by the alternating projections method applied to solve \eqref{Eqn:OptimizationMultiSet1}. We present the resulting procedure in Algorithm \ref{Algo:GLA}. \looseness=-1


Algorithm \ref{Algo:GLA} converges to a fixed-point, which is defined in the following sense. Let $\{ \boldsymbol{z}^k\}_{k \in \mathbb{N}}$ be a sequence generated by the algorithm with $\boldsymbol{z}^\infty$ denoting its accumulation point. Then, as shown in \cite[Prop. 1]{waldspurger2018phase}, the fixed-point iteration \eqref{Eqn:AltProj} converges, i.e.,
\begin{equation}
    \label{Eqn:FixedPointConvergence}
    \boldsymbol{z}^\infty = \boldsymbol{A}\boldsymbol{A}^\dagger \left( \mathrm{sgn}(\boldsymbol{z}^\infty) \odot \boldsymbol{b} \right), 
\end{equation}
under the assumption that $\boldsymbol{z}^\infty$ does not have any zero entries. If some entries of $\boldsymbol{z}^\infty$ are zeros, the result \eqref{Eqn:FixedPointConvergence} is rewritten more generally as $\boldsymbol{z}^\infty \in \boldsymbol{A}\boldsymbol{A}^\dagger P_Q(\boldsymbol{z}^\infty)$ (see \cite[Prop. 1]{waldspurger2018phase}). 

In general, the convergence of the fixed-point iteration does not imply convergence to a critical point.  Moreover, \eqref{Eqn:OptimizationMultiSet1} can be infeasible. 
The alternating projections method is also known to exhibit slow empirical convergence \cite{bauschke2004finding}. In the following sections, we address these issues.


\subsection{Solution Method Based on Projected Gradient Descent}
\label{Sec:ProjGrad}
\begin{algorithm}[t]
\caption{Projected gradient descent method to solve 
\eqref{Eqn:OptimizationMultiSetPenalty1}}
\begin{algorithmic}[1] 
\label{Algo:ProjGrad}
    \renewcommand{\algorithmicrequire}{\textbf{Input:}}
    \renewcommand{\algorithmicensure}{\textbf{Output:}}
    \REQUIRE $\boldsymbol{A}, \boldsymbol{b}, P_T$
\ENSURE $\boldsymbol{w}_T^\star$ 
    \STATE {Initialize} $\boldsymbol{{z}}^0 \in \mathbb{C}^{N_{\phi} \times 1}$ and choose $\Bar{\gamma} < 1$  
    \FOR{$k=1,2,...$}
    \STATE Update $\boldsymbol{z}^k$ according to \eqref{Eqn:ProjGradIterates_}
    \ENDFOR
    \STATE Return $\boldsymbol{w}_T^\star = \mathrm{nrm}\left( \boldsymbol{A}^\dagger \boldsymbol{z}^k \right)$
\end{algorithmic} 
\end{algorithm}
As elaborated in Section \ref{Sec:GriffinLim},
\eqref{Eqn:OptimizationMultiSet1} can be infeasible, i.e., $C \cap Q = \varnothing$.
A common technique to address this feasibility issue is relaxing $\delta_{C}(\boldsymbol{z})$ with the penalization term ${(\rho/2)\Vert \boldsymbol{A}\boldsymbol{\Tilde{w}}_T - \boldsymbol{z}\Vert^2}$ where $\rho > 0$ is a penalty parameter. 

This leads to the following unconstrained optimization problem:  
\begin{equation}
    \label{Eqn:OptimizationMultiSetPenalty1}
    \min_{\boldsymbol{\Tilde{w}}_T, \boldsymbol{z}} \left\{ \delta_{Q}(\boldsymbol{z}) + \frac{\rho}{2}\Vert \boldsymbol{A}\boldsymbol{\Tilde{w}}_T - \boldsymbol{z}\Vert^2\right\}.  
\end{equation}  
It is straightforward to verify that the alternating minimization method applied to \eqref{Eqn:OptimizationMultiSetPenalty1} again gives the same iterates expressed in \eqref{Eqn:FixedPoint2}. 

The minimization of \eqref{Eqn:OptimizationMultiSetPenalty1} with respect to $\boldsymbol{\Tilde{w}}_T$ reduces to a standard least-squares problem, whose optimal solution is given by $\boldsymbol{\Tilde{w}}_T = \boldsymbol{A}^\dagger \boldsymbol{z}$ \cite{horn2012matrix}. Then, \eqref{Eqn:OptimizationMultiSetPenalty1} can equivalently be written as
\begin{equation}
    \label{Eqn:OptimizationMultiSetPenalty2}
    \min_{\boldsymbol{z}} \left\{ \delta_{Q}(\boldsymbol{z}) + \frac{\rho}{2}\Vert \boldsymbol{M}\boldsymbol{z} \Vert^2\right\},  
\end{equation}
where $ \boldsymbol{M} \coloneqq  \boldsymbol{A}\boldsymbol{A}^\dagger - \boldsymbol{I}_{N_{\phi}}$. Unlike \eqref{Eqn:OptimizationMultiSet1}, the problem \eqref{Eqn:OptimizationMultiSetPenalty2} always has an optimal solution since a continuous function ($ \frac{\rho}{2}\Vert \boldsymbol{M}\boldsymbol{z} \Vert^2$) is minimized over a compact set ($Q$) \cite{luke2019optimization}. 

We adopt the projected gradient descent method \cite{beck2017first} to solve \eqref{Eqn:OptimizationMultiSetPenalty2}, with the following iterates: 
\begin{equation}
    \label{Eqn:ProjGradIterates}
    \boldsymbol{z}^k \in P_Q\left(\boldsymbol{z}^{k-1} - \frac{\rho}{\gamma} \boldsymbol{Y}\boldsymbol{z}^{k-1}\right),
\end{equation}
where 
\begin{equation}
    \label{Eqn:DefnB}
    \boldsymbol{Y} \coloneqq \boldsymbol{M}^H \boldsymbol{M} =  \boldsymbol{I}_{N_{\phi}} - \boldsymbol{A}\boldsymbol{A}^\dagger,
\end{equation}
and $1/\gamma$ denotes the step-size. Here, we define ${\Bar{\gamma} \coloneqq \rho / \gamma = 1}$ and by \eqref{Eqn:PqSgnRelationship}, we rewrite \eqref{Eqn:ProjGradIterates} as 
\begin{equation}
    \label{Eqn:ProjGradIterates_}
    \boldsymbol{z}^k  = \mathrm{sgn}\left(\boldsymbol{z}^{k-1} - \Bar{\gamma} \boldsymbol{Y}\boldsymbol{z}^{k-1}\right) \odot \boldsymbol{b} \in P_Q\left(\boldsymbol{z}^{k-1} - \Bar{\gamma} \boldsymbol{Y}\boldsymbol{z}^{k-1}\right).
\end{equation}
In the special case ${\Bar{\gamma} = 1}$, \eqref{Eqn:ProjGradIterates_} becomes
\begin{equation}
    \label{Eqn:GerchbergSaxton}
    \boldsymbol{z}^k \in P_Q\left(P_C(\boldsymbol{z}^{k-1})\right),
\end{equation}
which is the alternating projections update as in \eqref{Eqn:AlternatingProjections}, with the interchanged order of projections. 

We present the complete solution procedure in Algorithm \ref{Algo:ProjGrad}. As shown in Appendix \ref{Appendix:ConvergenceProof}, the algorithm is provably convergent, i.e., $\{ \boldsymbol{z}^k \}_{k \in \mathbb{N}}$ converges to a critical point of \eqref{Eqn:OptimizationMultiSetPenalty2}. 

\subsection{Solution Method Based on Accelerated Projected Gradient Descent with Adaptive Momentum Restart}
\label{Sec:AccProjGrad}
\begin{algorithm}[t]
\caption{Accelerated projected gradient descent with adaptive momentum restart to solve 
\eqref{Eqn:OptimizationMultiSetPenalty1}}
\begin{algorithmic}[1] 
\label{Algo:AccPGD}
    \renewcommand{\algorithmicrequire}{\textbf{Input:}}
    \renewcommand{\algorithmicensure}{\textbf{Output:}}
    \REQUIRE $\boldsymbol{A}, \boldsymbol{b}, P_T$ 
\ENSURE $\boldsymbol{{w}}_T^\star$ 
    \STATE {Initialize} $\boldsymbol{{z}}^0 = \boldsymbol{\Bar{z}}^1 \in \mathbb{C}^{N_{\phi} \times 1}$, and fix $\Bar{\gamma} < 1$
    \FOR{$k=1,2,...$}
    \STATE Update $\boldsymbol{z}^k$ according to \eqref{Eqn:ProjGradIterates_Momentum}
    \IF{\eqref{Eqn:RestartCondition} holds}
    \STATE Set $\boldsymbol{\Bar{z}}^{k+1} = \boldsymbol{{z}}^k$ 
    \ELSE
    \STATE {Update $\boldsymbol{\Bar{z}}^{k+1}$ according to \eqref{Eqn:ProjGradIterates_zbar}}
    \ENDIF
    \ENDFOR
    \STATE Return $\boldsymbol{w}_T^\star = \mathrm{nrm}\left( \boldsymbol{A}^\dagger \boldsymbol{z}^k \right)$
\end{algorithmic} 
\end{algorithm}
The projected gradient descent method presented in Section \ref{Sec:ProjGrad} can be accelerated via momentum-based schemes, which were originally developed for the convex problems and shown to achieve faster convergence rates \cite{beck2017first}. 
Although only limited theoretical results exist for their convergence in non-convex problems \cite{pock2016inertial, nenov2023accelerated}, empirical studies across a variety of applications report significant performance gains \cite{pauwels2017fienup, luke2019optimization, vu2023local}. \looseness=-1

To write the accelerated variant of Algorithm \ref{Algo:ProjGrad}, we replace the projected gradient iterates \eqref{Eqn:ProjGradIterates_} with 
\begin{equation}
    \label{Eqn:ProjGradIterates_Momentum}
    \boldsymbol{z}^k  = \mathrm{sgn}\left(\boldsymbol{\Bar{z}}^{k} - \Bar{\gamma} \boldsymbol{Y}\boldsymbol{\Bar{z}}^{k}\right) \odot \boldsymbol{b} \in P_Q\left(\boldsymbol{\Bar{z}}^{k} - \Bar{\gamma} \boldsymbol{Y}\boldsymbol{\Bar{z}}^{k}\right),
\end{equation}
where $\boldsymbol{\Bar{z}}^k$ denotes an auxiliary variable that can be updated at each iteration according to
\begin{equation}
    \label{Eqn:ProjGradIterates_zbar}
    \boldsymbol{\Bar{z}}^{k+1} = \boldsymbol{{z}}^k + \frac{k-1}{k+2}(\boldsymbol{{z}}^{k}-\boldsymbol{{z}}^{k-1}). 
\end{equation}

It is also common to incorporate restarting strategies to further enhance the empirical performance of accelerated projected gradient methods. Here, we adopt the gradient-based adaptive restart scheme from \cite{o2015adaptive}, i.e., we set $\boldsymbol{\Bar{z}}^{k+1} = \boldsymbol{z}^k$ whenever the following condition holds: 
\begin{equation}
    \label{Eqn:RestartCondition}
    \re\left\{(\boldsymbol{\Bar{z}}^k- \boldsymbol{{z}}^{k})^H (\boldsymbol{{z}}^{k}-\boldsymbol{{z}}^{k-1})\right\} > 0.
\end{equation}
The complete procedure is summarized in Algorithm~\ref{Algo:AccPGD}.
While not provably convergent as Algorithm~\ref{Algo:ProjGrad}, it demonstrates superior empirical performance, as shown in Section~\ref{Sec:NumericalResults}.

\subsection{Implementation Aspects}
\label{Sec:ImplementationAspects}
In this section, we provide some general remarks on the implementation aspects of all the algorithms presented herein. First, no specific initialization method for $\boldsymbol{z}^0$ has been prescribed so far. The theoretical convergence results presented for Algorithm~\ref{Algo:GLA} and Algorithm~\ref{Algo:ProjGrad} are global in the sense that they hold for an arbitrary initialization, with Algorithm~\ref{Algo:ProjGrad} providing a stronger convergence guarantee. However, in non-convex problems, initialization points affect the empirical performance. In Section~\ref{Sec:NumericalResults}, we set $\boldsymbol{z}^0 = P_C(\boldsymbol{b}) = \boldsymbol{A}\boldsymbol{A}^\dagger \boldsymbol{b}$ for the numerical analysis. The termination criterion for all algorithms can be set as
\begin{equation}
    \label{Eqn:TerminationCondition}
    {\Vert \boldsymbol{z}^{k} - \boldsymbol{z}^{k-1} \Vert}  \le \epsilon {\Vert \boldsymbol{z}^{k-1} \Vert}   , 
\end{equation}
where $\epsilon > 0$ is the termination tolerance, which we set to $\epsilon = 10^{-8}$. This criterion terminates the algorithm when the relative change between successive iterates becomes sufficiently small.

We also emphasize that none of the algorithms presented in this manuscript require parameter tuning, which is a practical advantage. Algorithm~\ref{Algo:GLA} is parameter-free, while Algorithms~\ref{Algo:ProjGrad} and~\ref{Algo:AccPGD} only require the step-size parameter $\Bar{\gamma}$, which should be chosen as large as possible while satisfying $\Bar{\gamma} < 1$. This can simply be achieved by setting $\Bar{\gamma} = 1 - \epsilon$.

In terms of computational complexity, the dominant operation for all algorithms is the evaluation of the operation $\boldsymbol{A}\boldsymbol{A}^\dagger\boldsymbol{u}$ for some $\boldsymbol{u}$ in each iteration, which has a worst-case complexity of $O(M_T N_{\phi})$, noting that $\boldsymbol{A}$ (and thus $\boldsymbol{A}^\dagger$) remains unchanged across iterations. However, a direct implementation of this operation is undesirable due to the numerical instability associated with the explicit computation of $\boldsymbol{A}^\dagger$. Instead, $\boldsymbol{A}^\dagger\boldsymbol{u}$ should be implemented in a numerically stable manner, for example, via the QR decomposition of $\boldsymbol{A}$~\cite{horn2012matrix}.

\section{Numerical Results}
\label{Sec:NumericalResults}

\begin{figure*}[t]%
    \centering
    \begin{subfigure}[t]{0.33\textwidth}
        \includegraphics[width=\textwidth]{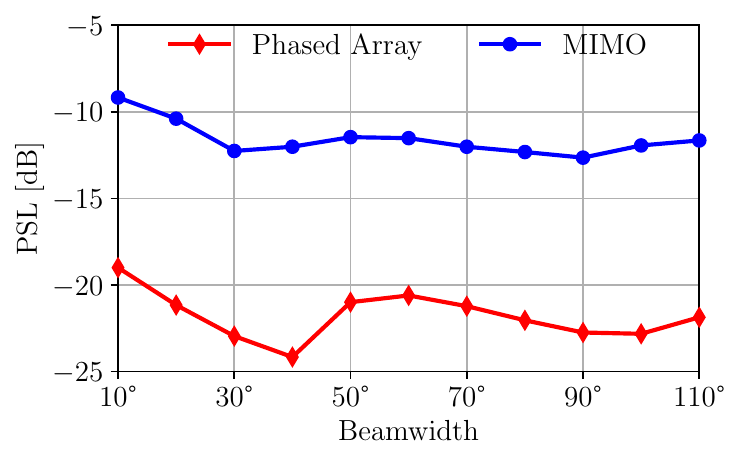}
        \subcaption{\Acrlongpl{psl}}
        \label{psl_bw}
    \end{subfigure}%
    \hfill
    \begin{subfigure}[t]{0.33\textwidth}
        \includegraphics[width=\textwidth]{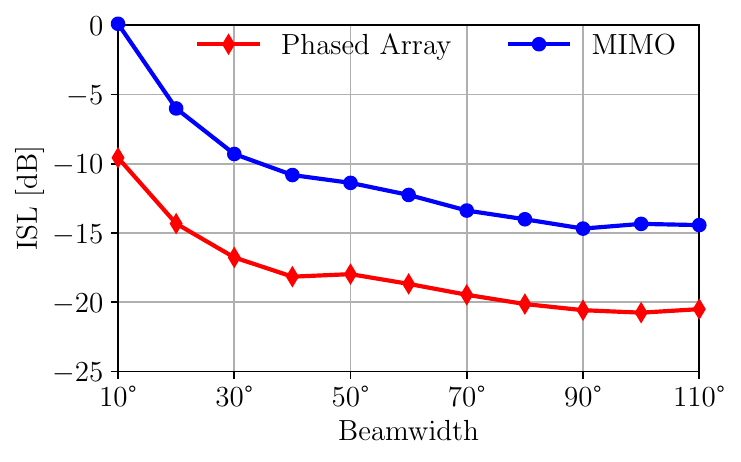}
        \subcaption{\Acrlongpl{isl}}
        \label{ismr_bw}
    \end{subfigure}%
    \hfill
    \begin{subfigure}[t]{0.33\textwidth}
        \includegraphics[width=\textwidth]{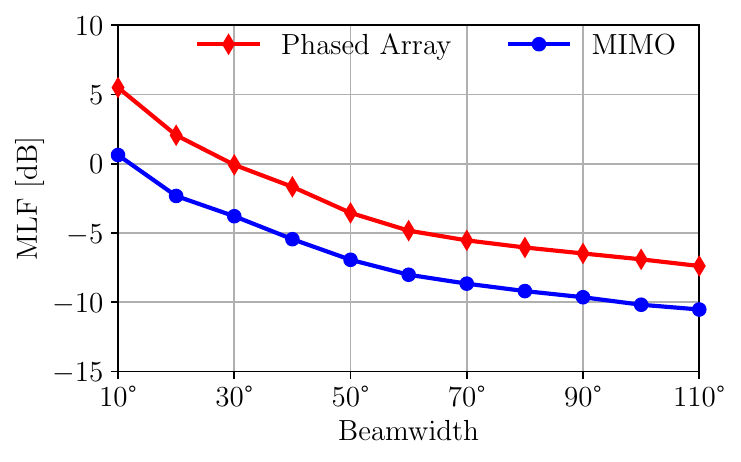}
        \subcaption{Mainlobe flatness values}
        \label{flatness_bw}
    \end{subfigure}%
    \caption{Synthesized \gls{tx} beampattern quality metrics as a function of desired beamwidths from $10^\circ$ to $110^\circ$ with $10^\circ$ steps, with the mainlobe centered at $\phi = 0^\circ$.}
    \label{fig:beampattern_quality_bw_phased}
\end{figure*}%

\begin{figure}[t]
    \centering
    \includegraphics[width=0.8\linewidth]{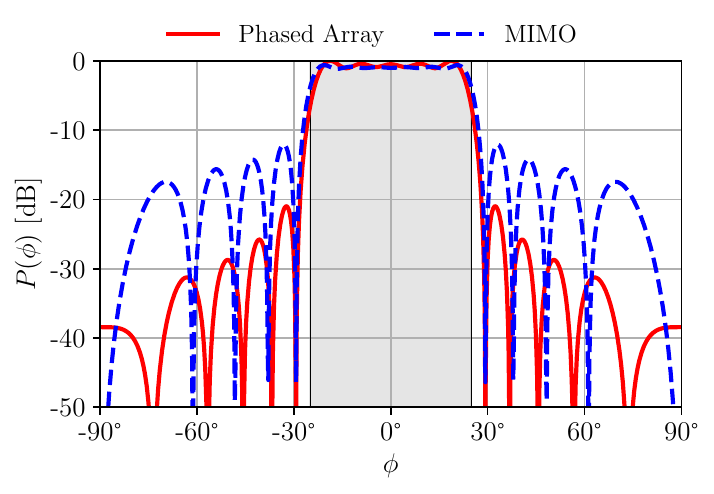} 
    \caption{Synthesized \gls{tx} beampatterns when the desired beamwidth is $50^\circ$ for a single mainlobe at $\phi = 0^{\circ}$.}
    \label{bp_bw50}
\end{figure}
\subsection{Considered Simulation Scenarios and Performance Metrics}
We consider a \gls{ula} with $\lambda/2$ inter-element spacing, for which the \gls{tx} array steering vector is:
\begin{equation}
    \label{Eqn:SteeringVector}
    \boldsymbol{a}_T(\phi)= [1~e^{j\pi \sin\phi}~...~e^{j(M_T-1) \pi \sin\phi }]^T.
\end{equation}
The grid size is set to $N_{\phi}=1024$ with grid points uniformly sampling the angular domain $[-90^\circ, 90^\circ]$, and the total power budget\footnote{For all algorithms proposed herein, $P_T$ only affects the final normalization, and any value of $P_T$ can be accommodated by a simple scaling of the solution.} is set to $P_T = 43$ dBm. 

In Section~\ref{Subsec:ChangingBW}, we analyze the quality of synthesized beampatterns for a single mainlobe under varying target beamwidths. In Section~\ref{Subsec:ChangingDOA}, we simulate a two-mainlobe scenario, where one mainlobe remains fixed while the central direction of the other one changes. In these two sections, we do not consider interference effects, i.e., we set ${\omega} = 1$ in \eqref{Eqn:DbarDefn}. We set the number of \gls{tx} antennas to $M_T = 16$ and present the \gls{mimo} radar results as a benchmark, obtained by formulating \eqref{Eqn:SDP_Formulation} in CVXPY~\cite{diamond2016cvxpy} and solving it with the modern general-purpose interior-point solver Clarabel~\cite{Clarabel_2024}. The proposed phased-array solution is obtained with Algorithm~\ref{Algo:ProjGrad}, as the empirical performances of Algorithms~\ref{Algo:GLA}, \ref{Algo:ProjGrad}, and~\ref{Algo:AccPGD} are very similar apart from their convergence speeds, which we discuss separately. The computation times are evaluated using a computer with Intel Core Ultra 5-135U CPU and 16 GB RAM. The initialization and stopping criteria of these algorithms are presented in Section \ref{Sec:ImplementationAspects}. 

In Section~\ref{Subsec:Jammer}, we demonstrate the interference suppression capability of the proposed method and evaluate its end-to-end performance in a monostatic radar direction-of-arrival estimation scenario. The conventional Bartlett beamformer \cite{van1988beamforming} is employed at the \gls{rx}, assuming identical \gls{tx} and \gls{rx} array configurations. Simulations are conducted for both $16$-antenna and $128$-antenna setups to illustrate the scalability of the proposed method to massive arrays.

For a quantitative analysis of the beampattern synthesis performance, we report several quality metrics. 
The level of power dissipation in unintended directions is commonly assessed using the \gls{psl}, which is evaluated relative to the maximum power in the mainlobe region \cite{aubry2016mimo}. In contrast, the efficiency of the synthesized beampattern in directing energy toward desired directions is quantified by the \gls{isl}, defined as
\begin{equation}
    \mathrm{ISL} \coloneqq \frac{\sum_{\bar{\phi}_i \in \Phi_s} P(\bar{\phi}_i)}{\sum_{\bar{\phi}_j \in \Phi_m} P(\bar{\phi}_j)}.
\end{equation}

For the \gls{mlf}, as the performance evaluation metric, we consider the standard deviation within the mainlobe region, normalized by $P_T$, which we define as: 
\begin{equation}
    \mathrm{MLF} \coloneqq \frac{1}{P_T}\sqrt{\frac{\sum_{\bar{\phi}_i \in \Phi_m} \left( P(\bar{\phi}_i) - \Bar{P} \right)^2 }{ |\Phi_m|}},
\end{equation}
where $\Bar{P}$ is the mean of $P(\bar{\phi}_i)$ computed over all $\bar{\phi}_i \in \Phi_m$, and $|\Phi_m|$ denotes the cardinality of $\Phi_m$, i.e., the number of grid points corresponding to the mainlobe region.  

In the two-mainlobe scenario considered in Section~\ref{Subsec:ChangingDOA}, to evaluate the balance of \gls{tx} power allocated between the two mainlobes, we introduce the \gls{mmpar} metric, defined as
\begin{equation}
    \label{Eqn:MMPAR}
    \mathrm{MMPAR} \coloneqq 
    \frac{
    \min \left\{
    \sum_{\bar{\phi}_i \in \Phi_m^{(1)}} P(\bar{\phi}_i), \;
    \sum_{\bar{\phi}_j \in \Phi_m^{(2)}} P(\bar{\phi}_j)
    \right\}
    }{
    \sum_{\bar{\phi}_i \in \Phi_m} P(\bar{\phi}_i)
    },
\end{equation}
where $\Phi_m^{(1)}$ and $\Phi_m^{(2)}$ denote the sets of angular grid points corresponding to the two distinct mainlobe regions, such that $\Phi_m^{(1)} \cup \Phi_m^{(2)} = \Phi_m$ and $\Phi_m^{(1)} \cap \Phi_m^{(2)} = \varnothing$. This metric measures how evenly the total power is distributed between the two mainlobes. A value close to $0.5$ indicates that both lobes receive nearly equal power, while smaller values indicate that one lobe dominates.

For visualization, we additionally present the normalized \gls{tx} beampatterns for selected scenarios, where the normalization is performed using the largest \gls{tx} beampattern value computed among the compared methods such that $0$ dB corresponds to the global peak.

\begin{figure*}%
    \centering
    \begin{subfigure}[t]{\figWidth cm}
        \includegraphics[width=\figWidth cm]{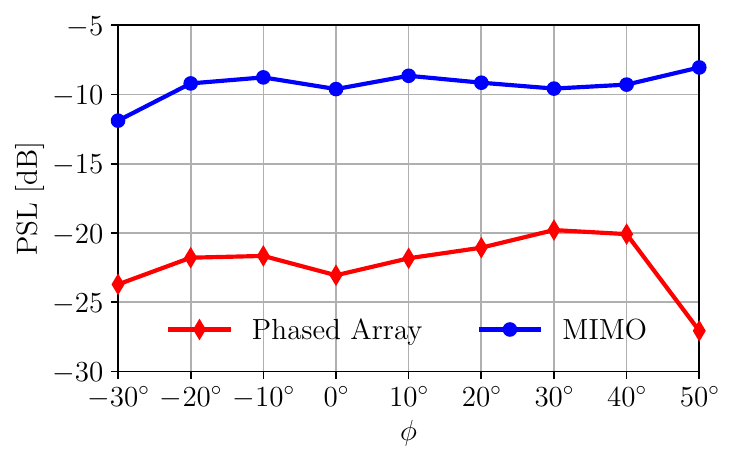}
        \subcaption{\Acrlongpl{psl}}
        \label{psl_doa}
    \end{subfigure}%
    \begin{subfigure}[t]{\figWidth cm}
        \includegraphics[width=\figWidth cm]{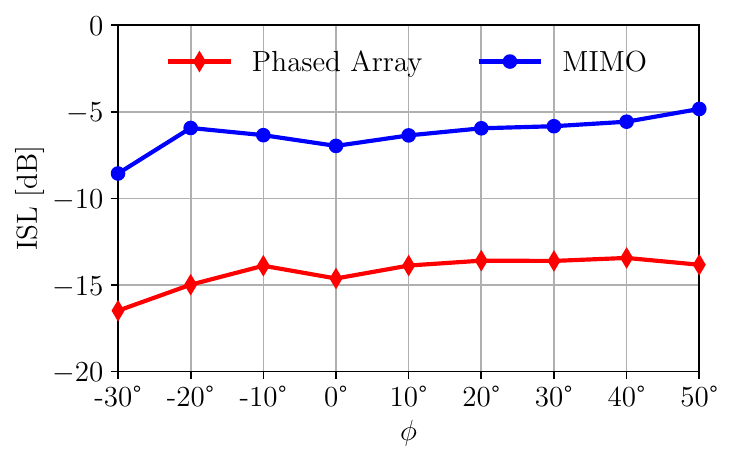}
        \subcaption{\Acrlongpl{isl}}
        \label{ismr_doa}
    \end{subfigure}%
    \begin{subfigure}[t]{\figWidth cm}
        \includegraphics[width=\figWidth cm]{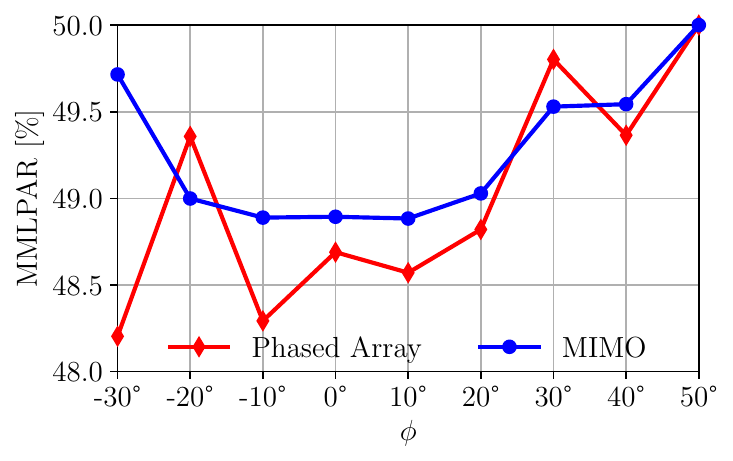}
        \subcaption{\Acrlong{mmpar}s\looseness=-1}
        \label{power_doa}
    \end{subfigure}%
    \caption{Synthesized \gls{tx} beampattern quality metrics for a dual-mainlobe scenario, with one mainlobe fixed at $-50^\circ$ and the other scanning from $-30^\circ$ to $50^\circ$ with $10^\circ$ steps.}
    \label{fig:beampattern_quality_doa_phased}
\end{figure*}%
\subsection{Impact of Mainlobe Beamwidth}
\label{Subsec:ChangingBW}
We conduct numerical experiments with a single beam, by varying the desired beamwidth from $10^{\circ}$ to $110^{\circ}$ in increments of $10^{\circ}$, while keeping the mainlobe center fixed at $0^{\circ}$. The ability to support a flexible mainlobe beamwidths is essential in beamforming design, e.g., short-range radars often require broader mainlobes, whereas long-range radars benefit from narrower beams.

Fig.~\ref{psl_bw} shows the \gls{psl} as a function of the beamwidth. As we observe, the proposed phased-array radar solution consistently achieves lower \gls{psl} values than the \gls{mimo} radar solution. On average, the \gls{mimo} radar yields a \gls{psl} of $-11.6$~dB, whereas the phased-array radar achieves $-21.8$~dB. This demonstrates a substantial sidelobe suppression advantage of the phased-array radar enabled by the proposed method.

Fig.~\ref{ismr_bw} reports the \gls{isl} as a function of beamwidth. Again, the proposed phased-array radar solution outperforms the \gls{mimo} radar solution. The \gls{mimo} radar attains an average \gls{isl} of $-11$~dB, compared to $-17.9$~dB for the phased-array solution. Figs.~\ref{psl_bw} and~\ref{ismr_bw} together highlight that the proposed phased-array solution allocates \gls{tx} power more efficiently compared to its \gls{mimo} radar counterpart.

Fig.~\ref{flatness_bw} examines the mainlobe flatness, measured as the standard deviation of the normalized \gls{tx} beampattern within the mainlobe region. In this metric, the \gls{mimo} radar outperforms the phased-array solution, achieving a flatter mainlobe response. 
This reflects a trade-off between  sidelobe suppression and \acrlong{mlf} between the \gls{mimo} and phased-array radars.

Fig.~\ref{bp_bw50} shows the synthesized beampatterns for the two radar types, for a desired beamwidth of $50^{\circ}$, with the shaded gray region indicating the target mainlobe region\footnote{The same scenario as illustrated in Fig. \ref{polar_bp}.}. It is observed that the proposed phased-array solution achieves substantially lower sidelobes, albeit at the cost of greater variation across the mainlobe. In contrast, the \gls{mimo} radar solution yields a flatter mainlobe but exhibits higher sidelobes. These observations visually confirm the trends quantitatively reported in Fig.~\ref{fig:beampattern_quality_bw_phased}.

In Table~\ref{tab:iterations_bw}, we report the iteration numbers at which the algorithms terminate according to the criterion in~\eqref{Eqn:TerminationCondition}, for the scenarios considered in this section. Algorithms~\ref{Algo:GLA} and~\ref{Algo:ProjGrad} exhibit nearly identical convergence behavior.
In contrast, Algorithm~\ref{Algo:AccPGD} converges significantly faster, typically terminating by an order of magnitude earlier than the other two algorithms.

By considering the computation times averaged across all cases, we observe that Algorithms~\ref{Algo:GLA} and~\ref{Algo:ProjGrad} both require approximately 0.25 s to converge, while Algorithm~\ref{Algo:AccPGD} completes in about 0.03 s. In contrast, the \gls{mimo} radar solution is considerably slower, taking around 113 s to obtain a solution. This difference reflects the fact that the \gls{mimo} formulation optimizes $M_T^2$ variables, compared to only $M_T$ variables in the phased-array case.

\begin{table}[t]
\centering
\caption{Number of iterations for convergence for different mainlobe beamwidths from $10^\circ$ to $110^\circ$.}
\label{tab:iterations_bw}
\setlength{\tabcolsep}{3pt} 
\begin{tabular}{c|ccccccccccc}
\hline
\textbf{} & $10^\circ$ & $20^\circ$ & $30^\circ$ & $40^\circ$ & $50^\circ$ & $60^\circ$ & $70^\circ$ & $80^\circ$ & $90^\circ$ & $100^\circ$ & $110^\circ$ \\
\hline
Alg. \ref{Algo:GLA}     & 30 & 150 & 441 & 5804 & 716 & 526 & 605 & 640 & 717 & 513 & 369 \\
Alg. \ref{Algo:ProjGrad} & 31 & 151 & 444 & 5828 & 718 & 527 & 607 & 642 & 720 & 515 & 370 \\
Alg. \ref{Algo:AccPGD}   & 26 & 36  & 62  & 116  & 66  & 56  & 61  & 63  & 67  & 68  & 48    \\
\hline
\end{tabular}
\end{table}

\begin{figure}[t]
    \centering
    \includegraphics[width=0.8\linewidth]{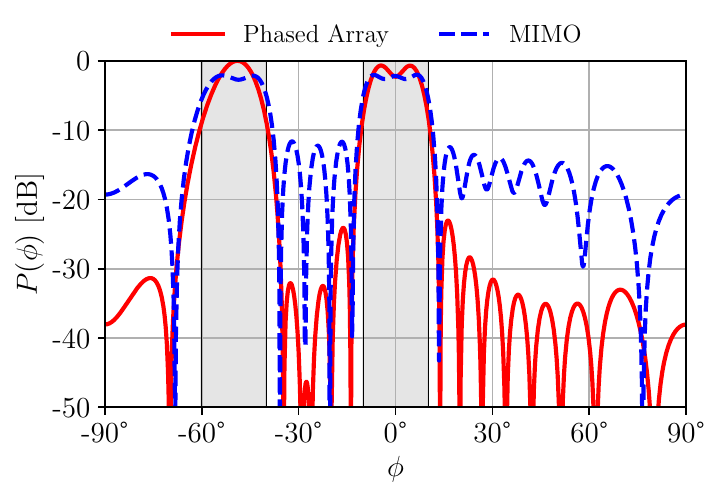} 
   \caption{Synthesized \gls{tx} beampatterns with two mainlobes centered at $\phi = 0^{\circ}$ and $\phi = -50^{\circ}$, each having a desired beamwidth of $20^\circ$.}
    \label{bp_doa0}
\end{figure}

\begin{table}[t]
\centering
\caption{Number of iterations for convergence for different mainlobe directions from $-30^\circ$ to $50^\circ$.}
\label{tab:iterations_doa}
\setlength{\tabcolsep}{3pt} 
\begin{tabular}{c|ccccccccc}
\hline
\textbf{} & $-30^\circ$ & $-20^\circ$ & $-10^\circ$ & $0^\circ$ & $10^\circ$ & $20^\circ$ & $30^\circ$ & $40^\circ$ & $50^\circ$ \\
\hline
Alg. \ref{Algo:GLA}     & 540  & 7372 & 1006 & 6151 & 1784 & 1574 & 2214 & 1394 & 9277 \\
Alg. \ref{Algo:ProjGrad} & 542  & 7399 & 1007 & 6162 & 1786 & 1575 & 2218 & 1396 & 9283 \\
Alg. \ref{Algo:AccPGD}   & 51   & 162  & 100  & 160  & 106  & 101  & 114  & 97   & 141  \\
\hline
\end{tabular}
\end{table}

\subsection{Impact of the Central Mainlobe Direction}
\label{Subsec:ChangingDOA}
We analyze the quality of the synthesized beampatterns by simulating two mainlobes, i.e., one fixed at $\phi = -50^{\circ}$ and the other scanning from $\phi = -30^{\circ}$ to $\phi = 50^{\circ}$ with $10^{\circ}$ steps. The desired beamwidth is set to $20^{\circ}$ for both mainlobe regions.

Fig.~\ref{psl_doa} shows the variation of the \gls{psl} with respect to the scanning mainlobe direction. The proposed phased-array radar solution consistently achieves lower \gls{psl} values than the \gls{mimo} radar in all cases. On average, the \gls{mimo} radar attains a \gls{psl} of $-9.4$~dB, whereas the phased-array radar achieves $-22.2$~dB, again highlighting a significant sidelobe suppression advantage enabled by the proposed approach.

Fig.~\ref{ismr_doa} presents the \gls{isl} as a function of the scanning direction. Once again, the phased-array radar demonstrates superior performance, achieving an average \gls{isl} of $-14.3$~dB compared to $-6.3$~dB for the \gls{mimo} radar. Figs.~\ref{psl_doa} and~\ref{ismr_doa} together confirm that the proposed method yields more efficient \gls{tx} power allocation also in the two-mainobe scenario.

Fig.~\ref{power_doa} analyzes the power distribution between the two mainlobes using the \gls{mmpar} metric \eqref{Eqn:MMPAR}. An ideal uniform illumination corresponds to \gls{mmpar} of $50 \%$. Both the \gls{mimo} and phased-array radars achieve near-uniform power allocation, with average \gls{mmpar} values of $49.3 \%$ and $49.0 \%$, respectively, indicating approximately the same performance.

Fig.~\ref{bp_doa0} depicts the synthesized beampatterns when the scanning mainlobe is centered at $\phi = 0^{\circ}$, with the shaded gray regions indicating the target mainlobe regions. The phased-array radar exhibits substantially lower sidelobes and an almost uniform radiation pattern across both mainlobe regions, consistent with the quantitative results reported in Fig.~\ref{fig:beampattern_quality_doa_phased}.

In Table~\ref{tab:iterations_doa}, we report the iteration numbers at which the algorithms terminate for the scenarios considered in this section. As observed earlier, Algorithm~\ref{Algo:AccPGD} achieves convergence with significantly fewer iterations. In terms of computation times, Algorithms~\ref{Algo:GLA} and~\ref{Algo:ProjGrad} both require approximately 1~s to converge on average, whereas Algorithm~\ref{Algo:AccPGD} completes in about 0.04 s. The \gls{mimo} radar beamforming design algorithm is considerably slower, taking around 50 s to obtain a solution.

\begin{figure}[t]
    \centering
    \includegraphics[width=0.85\linewidth]{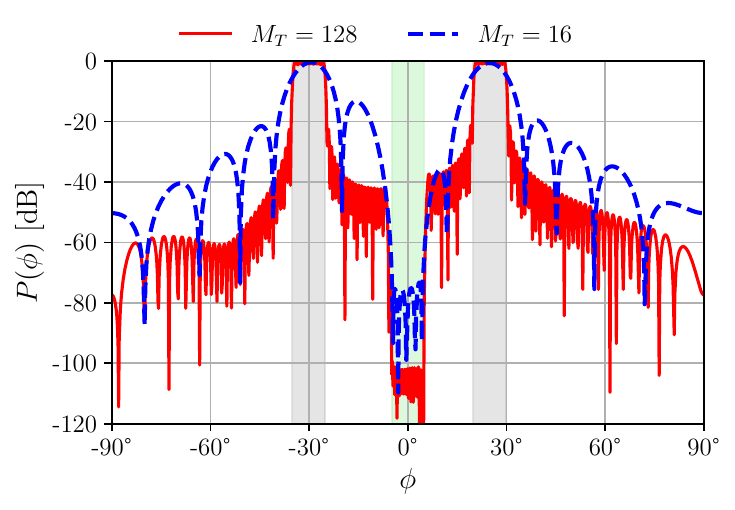} 
   \caption{Synthesized \gls{tx} beampatterns with radiation nulling for the interference region $[-5^\circ, 5^\circ]$ for $M_T = 16$ and $M_T = 128$.}
    \label{jammer_tx_overlay}
\end{figure}

\subsection{The Effect of Interference Suppression}
\label{Subsec:Jammer}
In the results presented so far, interference effects are not considered, i.e., corresponding to the case with $\omega = 1$ in \eqref{Eqn:DbarDefn}. Here, we evaluate the interference suppression performance of the proposed method. We simulate a two-mainlobe scenario with the mainlobes centered at $\phi = -30^\circ$ and $\phi = 25^\circ$, both with a target beamwidth of $10^\circ$. The interference is modeled to span the angular region from $\phi = -5^\circ$ to $\phi = 5^\circ$. To enable interference suppression, we set $\omega = 10^3$ in this case. 

Fig. \ref{jammer_tx_overlay} shows \gls{tx} beampatterns obtained for $M_T = 16$ and $M_T = 128$. As expected, the beampattern quality improves with the larger array. For $M_T = 16$, an interference suppression of approximately $-80$ dB is achieved, whereas for $M_T = 128$, the suppression exceeds $-100$ dB. These results confirm the effectiveness of the method in mitigating interference. \looseness=-1

Regarding the computation times, in this case Algorithm \ref{Algo:AccPGD} requires around 0.03 s for $M_T = 16$ and 0.7 s for $M_T = 128$. Despite the substantial increase in the antenna count, the method is still able to generate the output under 1 s. \looseness=-1

To demonstrate the impact of interference suppression in a direction-of-arrival estimation scenario, we consider two radar targets located at $\phi_1 = -32^\circ$ and $\phi_2 = 27^\circ$, whose reflection coefficients are modeled as $q(\phi_1) = e^{j\psi_1}$ and $q(\phi_2) = e^{j\psi_2}$, respectively. A strong interferer is assumed to be located at $\phi_3 = 2^\circ$ with a reflection coefficient of $q(\phi_3) = 1000 e^{j\psi_3}$, corresponding to a \acrlong{sir} of $-60$~dB. Random phase shifts are incorporated into the reflection coefficients of both the targets and the interferer, with $\psi_i \sim \mathcal{U}[0, 2\pi]$, where $\mathcal{U}$ denotes the uniform distribution. Note that the radar targets lie within the mainlobe regions of the \gls{tx} beampattern illustrated in Fig.~\ref{jammer_tx_overlay}, whereas the jammer is positioned within its suppressed region. In this case, following \eqref{Eqn:RadarRx}, the \gls{rx} signal is modeled as the superposition of the echoes from both targets and the interferer, i.e.,
\begin{equation}
\label{Eqn:RadarRx2}
\boldsymbol{y}_R(t) \coloneqq \sum_{i=1}^{3}q(\phi_i)\boldsymbol{a}^*_R(\phi_i)\boldsymbol{a}_T^H(\phi_i)\boldsymbol{x}(t) + \boldsymbol{n}_R(t), 
\end{equation}
where $\boldsymbol{a}_R(\phi_i) = \boldsymbol{a}_T(\phi_i)$, $1 \le i \le 3$, since we consider a monostatic radar direction-of-arrival estimation scenario. 

\begin{figure}[t]
    \centering
    \includegraphics[width=0.8\linewidth]{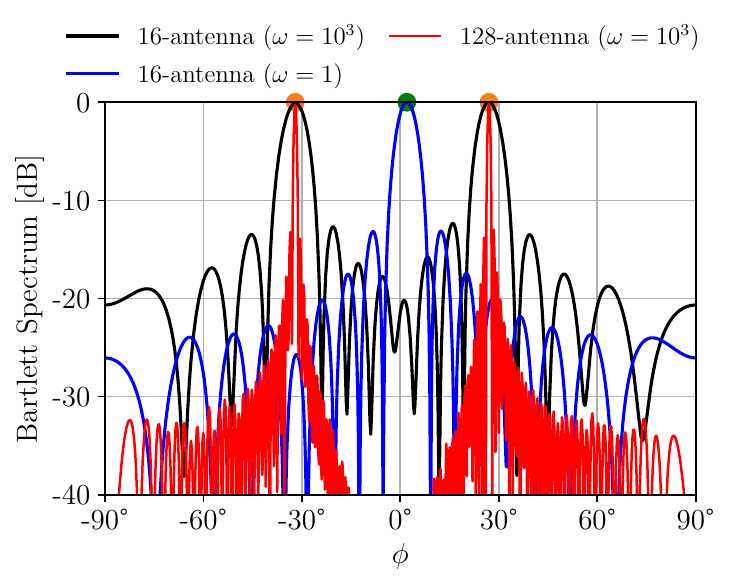} 
    \caption{Normalized Bartlett spectra illustrating the effect of interference suppression on angle estimation performance.}
    \label{jammer_bartlett}
\end{figure}

Fig.~\ref{jammer_bartlett} shows the normalized angle estimation spectra obtained using the Bartlett beamformer \cite{van1988beamforming} at an \gls{snr} of $10$~dB, where the \gls{snr} is defined as $\mathrm{SNR} = P_T/\sigma_n^2$.
The interference and target directions are indicated by green and orange dots, respectively. With interference suppression ($\omega = 10^3$), the Bartlett spectrum exhibits clear peaks at the true target locations for both array configurations, enabling accurate angle estimation. The 128-antenna setup produces sharper peaks, indicating higher angular resolution, as expected.

In contrast, for the 16-antenna case without interference suppression ($\omega = 1$), the strong interferer makes the two targets indistinguishable. For visual clarity, the 128-antenna scenario without interference suppression is not shown, as the same effect occurs. This result highlights the effectiveness of the proposed \gls{tx} beamformer design in maintaining estimation accuracy under strong interference. 

\section{Conclusions}
\label{Sec:Conclusions}
In this work, we addressed the phased-array \gls{tx} beamformer design for \gls{bpm} and interference suppression. We formulated a non-convex optimization framework that jointly incorporates these two objectives. We then proposed a solution method based on alternating minimization, which splits the problem into two subproblems that admit closed-form solutions, one of which recovers the classical maximum-gain and null-steering phased-array beamformers as special cases. 

We further reformulated the beamformer design as a non-convex set feasibility problem, enabling the development of a projected gradient descent algorithm with guaranteed convergence despite the non-convexity of the formulation. Finally, we designed an accelerated variant of this method to further improve empirical convergence. 
Extensive numerical results show that the proposed approach, despite having fewer degrees of freedom, attains lower \gls{psl} and \gls{isl} values, thereby indicating superior \gls{tx} beampattern synthesis performance, while providing significantly higher computational efficiency compared to a widely adopted \gls{mimo} radar design.

\appendices
\section{}
\label{Appendix:SpecialCases}
For the null-steering beamformer \eqref{Eqn:PhasedBF_nullSteer}, we set $\Phi_g = \Phi_J  \cup \{ \Bar{\phi} \}$ and, without loss of generality, ${\omega} = 1$ in \eqref{Eqn:DbarDefn}, resulting in ${\boldsymbol{\Bar{D}}=[\boldsymbol{D}_J ~ \boldsymbol{a}_T(\Bar{\phi})]}$ and ${\boldsymbol{b}=[\boldsymbol{0}^T ~1]^T}$. We note that the columns of $\boldsymbol{\Bar{D}}$ can arbitrarily be permuted as long as $\boldsymbol{b}^T$ is permuted in the same way. For this special case, \eqref{Eqn:Eta_x_update} becomes 
\begin{align}
    \label{Eqn:Wt_NullSteering1}
    \boldsymbol{w}_T^\diamond = \Bar{\eta}^\diamond (\boldsymbol{\Bar{D}}^H)^\dagger 
    \begin{bmatrix} \boldsymbol{0} \\ 1 
    \end{bmatrix}.
\end{align}
Since $\boldsymbol{\Bar{D}}$ is a tall matrix in this case, the pseudoinverse term can be written as
\begin{equation}
\label{Eqn:AppPseudoInv}
    (\boldsymbol{\Bar{D}}^H)^\dagger
    = \boldsymbol{\Bar{D}} (\boldsymbol{\Bar{D}}^H \boldsymbol{\Bar{D}})^{-1}
 = \begin{bmatrix}
        \boldsymbol{D}_J & 
        \boldsymbol{a}_T(\Bar{\phi}) 
    \end{bmatrix}  \begin{bmatrix}
        \boldsymbol{C}_1 & \boldsymbol{c}_2 \\ 
        \boldsymbol{C}_3 & {c}_4
    \end{bmatrix},
\end{equation}
where we introduce the following notation
\begin{align}
    \begin{bmatrix}
        \boldsymbol{C}_1 & \boldsymbol{c}_2 \\ 
        \boldsymbol{C}_3 & {c}_4
    \end{bmatrix}
    &\coloneqq  
    (\boldsymbol{\Bar{D}}^H \boldsymbol{\Bar{D}})^{-1} \\
    &=
    \begin{bmatrix}
        \boldsymbol{D}_J^H\boldsymbol{D}_J & \boldsymbol{D}_J^H\boldsymbol{a}_T(\Bar{\phi}) \\ 
        \boldsymbol{a}_T^H(\Bar{\phi})\boldsymbol{D}_J & \Vert \boldsymbol{a}_T(\Bar{\phi})\Vert^2
    \end{bmatrix}^{-1}.
\end{align}
By substituting \eqref{Eqn:AppPseudoInv} into \eqref{Eqn:Wt_NullSteering1}, we obtain
\begin{align}
    \label{Eqn:Appw_t}
    \boldsymbol{w}_T^\diamond &= \Bar{\eta}^\diamond \begin{bmatrix}
        \boldsymbol{D}_J & 
        \boldsymbol{a}_T(\Bar{\phi}) 
    \end{bmatrix}
    \begin{bmatrix}
        \boldsymbol{C}_1 & \boldsymbol{c}_2 \\ 
        \boldsymbol{C}_3 & {c}_4
    \end{bmatrix} \begin{bmatrix}
        \boldsymbol{0}\\ 
        1 
    \end{bmatrix} \nonumber \\
    &= \Bar{\eta}^\diamond \begin{bmatrix}
        \boldsymbol{D}_J & 
        \boldsymbol{a}_T(\Bar{\phi}) \end{bmatrix} \begin{bmatrix}
        \boldsymbol{c}_2\\ 
        c_4 
    \end{bmatrix} \nonumber \\ 
    &=\Bar{\eta}^\diamond  \left( \boldsymbol{D}_J \boldsymbol{c}_2 + \boldsymbol{a}_T(\Bar{\phi}) c_4 \right). 
\end{align}

From the classic result on block matrix inversion \cite[Ch.~0.7.3]{horn2012matrix}, we have
\begin{align}
    \label{Eqn:Appc2}
    \boldsymbol{c}_2 &= -(\boldsymbol{D}_J^H\boldsymbol{D}_J)^{-1} {\boldsymbol{D}^H_J}\boldsymbol{a}_T(\Bar{\phi})c_4,
    \\
    \label{Eqn:Appc4}
    c_4 &= \left( \Vert \boldsymbol{a}_T(\Bar{\phi})\Vert^2-\boldsymbol{a}_T^H(\Bar{\phi})\boldsymbol{D}_J (\boldsymbol{D}_J^H\boldsymbol{D}_J)^{-1}\boldsymbol{D}_J^H\boldsymbol{a}_T(\Bar{\phi}) \right)^{-1}.
\end{align}
By substituting \eqref{Eqn:Appc2} and \eqref{Eqn:Appc4} into \eqref{Eqn:Appw_t}, we obtain 
\begin{equation}
    \label{Eqn:Wt_NullSteering3}
    \boldsymbol{w}_T^\diamond = \Bar{\eta}^\diamond  c_4 \left( \boldsymbol{I}_{M_T} -  \boldsymbol{D}_J(\boldsymbol{D}_J^H\boldsymbol{D}_J)^{-1} {\boldsymbol{D}^H_J} \right) \boldsymbol{a}_T(\Bar{\phi}),  
\end{equation}
which is equivalent to \eqref{Eqn:PhasedBF_nullSteer}, since $\Bar{\eta}^\diamond $ accounts for power normalization and ensures that $\Vert \boldsymbol{w}_T^\diamond \Vert^2 = P_T$. 

On the other hand, when $\Phi_g = \{ \Bar{\phi} \}$, we have ${\boldsymbol{\Bar{D}}=[\boldsymbol{a}_T(\Bar{\phi})]}$ and ${\boldsymbol{b}=1}$. In this case, the solution in \eqref{Eqn:Eta_x_update} straightforwardly reduces to the conventional maximum-gain phased-array beamformer given in \eqref{Eqn:PhasedBF_maxSNR}.

\section{}
\label{Appendix:ConvergenceProof}

The convergence of Algorithm \ref{Algo:ProjGrad} follows from the following theorem, which we adopt from \cite[Thm. 5.3]{attouch2013convergence} and \cite[Prop. 3]{bolte2014proximal}: 
\begin{theorem}[The Global Convergence of Non-convex Projected Gradient Descent]
\label{Thm:ConvergencePG}
Let $G$ be a continuously differentiable function with an $L$-Lipschitz continuous gradient, and $\mathcal{X}$ be a nonempty and closed set. Let $\{ \boldsymbol{u}^k \}_{k \in \mathbb{N}}$ be a bounded sequence generated by 
\begin{equation}
    \boldsymbol{u}^k \in P_\mathcal{X}\left(\boldsymbol{u}^{k-1} - {\frac{1}{\gamma}} \nabla G(\boldsymbol{u}^{k-1})\right). 
\end{equation}
If $\delta_\mathcal{X} + G$ is a \gls{kl} function and ${\gamma > L}$, then $\{ \boldsymbol{u}^k \}_{k \in \mathbb{N}}$ has a finite length, i.e., ${\sum_{k=1}^\infty \Vert \boldsymbol{u}^{k+1} - \boldsymbol{u}^{k} \Vert < \infty}$, and it converges to a critical point (defined based on the limiting subdifferential) of $\delta_\mathcal{X} + G$. 
\end{theorem}

To establish the convergence of Algorithm \ref{Algo:ProjGrad}, it suffices to verify that the assumptions of Theorem \ref{Thm:ConvergencePG} apply. 
First, we define the objective function of \eqref{Eqn:OptimizationMultiSetPenalty2} as
\begin{equation}
    \label{Eqn:Fdefn}
    F(\boldsymbol{z}) \coloneqq \delta_{Q}(\boldsymbol{z}) + H(\boldsymbol{z}),
\end{equation}
where $H(\boldsymbol{z}) \coloneqq (\rho/2) \Vert \boldsymbol{Mz} \Vert^2$.

By Theorem \ref{Thm:ConvergencePG}, $F$ is required be a \gls{kl} function (see \cite[Def. 2.4]{attouch2013convergence} for the definition). 
It is known that proper lower semi-continuous and semi-algebraic functions are \gls{kl} functions \cite[Thm. 6.1]{bolte2018first}. 
We first start with the proper lower semi-continuity assumption. $\delta_Q$ in \eqref{Eqn:OptimizationMultiSetPenalty2} is proper and lower semi-continuous since the set $Q$ is nonempty and closed (the indicator function of a non-empty closed set is proper and lower semi-continuous), and $H(\boldsymbol{z})$ is continuous and finite-valued. Then, $F$ is a proper lower semi-continuous function. 

We now need to show that $F$ is also semi-algebraic. Semi-algebraic sets are defined as the sets that can be constructed by finite number of polynomial inequalities \cite{luke2019optimization} (see \cite[Def. 3.1]{attouch2013convergence} for the formal definition). The set $Q$ has been shown to be semi-algebraic in several studies, e.g., see \cite{hesse2015proximal, luke2019optimization}. Moreover, it is known that the indicator function of a semi-algebraic set is a semi-algebraic function \cite[Sec. 4.3]{attouch2010proximal}, i.e., $\delta_Q$ in \eqref{Eqn:Fdefn} is semi-algebraic. Furthermore, since $H$ is polynomial, it is also semi-algebraic. Finally, the sum of two semi-algebraic functions is semi-algebraic \cite{attouch2010proximal}, i.e., $F$ is semi-algebraic. Therefore, $F$ is a \gls{kl} function. 

Furthermore, the sequence $\{ \boldsymbol{z}^k \}_{k \in \mathbb{N}}$ generated by Algorithm \ref{Algo:ProjGrad} is required to be bounded. This is already satisfied here since all iterates of the algorithm stay within the set $Q$ \eqref{Eqn:SetQ}, i.e., bounded by definition, making the sequence $\{ \boldsymbol{z}^k \}_{k \in \mathbb{N}}$ automatically bounded (see \cite[Remark 3.1]{hesse2015proximal}). 

Finally, the gradient of $H$ is required to be Lipschitz continuous, i.e.,
the following must hold $\forall \boldsymbol{u}, \boldsymbol{v}$:
\begin{equation}
    \Vert \nabla H(\boldsymbol{u}) - \nabla H(\boldsymbol{v}) \Vert \le L \Vert \boldsymbol{u} - \boldsymbol{v} \Vert.
\end{equation}
Since $\nabla H(\boldsymbol{u}) = \rho \boldsymbol{M}^H \boldsymbol{Mu}$, we have 
\begin{equation}
    \left\Vert \rho \boldsymbol{M}^H \boldsymbol{Mu} - \rho \boldsymbol{M}^H \boldsymbol{Mv} \right\Vert \le \rho \lambda_\mathrm{max}(\boldsymbol{M}^H \boldsymbol{M}) \Vert \boldsymbol{u} - \boldsymbol{v} \Vert,
\end{equation}
which follows from the Cauchy-Schwarz inequality \cite{horn2012matrix}. In our case, $\lambda_\mathrm{max}(\boldsymbol{M}^H \boldsymbol{M})=1$ (see \eqref{Eqn:DefnB}). This verifies that $\nabla{H}$ is $L$-Lipschitz continuous with $L = \rho$. 

Then, by Theorem \ref{Thm:ConvergencePG}, we conclude that the projected gradient sequence generated according to \eqref{Eqn:ProjGradIterates} is provably convergent for $\gamma > \rho$. In Algorithm \ref{Algo:ProjGrad}, we write the equivalent condition on the step-size as  $\Bar{\gamma} = \rho / \gamma < 1$. Therefore, we conclude that Algorithm \ref{Algo:ProjGrad} is provably convergent, i.e., the sequence $\{ \boldsymbol{z}^k \}_{k \in \mathbb{N}}$ generated by the algorithm has a finite length and it converges to a critical point of $F$.

\bibliographystyle{IEEEtran}
\bibliography{ieeeTran_bibliography_isac}

\end{document}